\documentclass[showkeys]{revtex4}
\usepackage{graphicx}
\usepackage{dcolumn}
\usepackage{amsmath}
\usepackage{color}
\usepackage{epstopdf}
\usepackage{float}
\usepackage{tikz}
\usepackage{graphicx}
\usepackage{subfigure}
\usepackage[utf8x]{inputenc}
\definecolor{coolblack}{rgb}{0.0, 0.18, 0.39}
\definecolor{darkred}{rgb}{0.5,0,0}
\definecolor{darkgreen}{rgb}{0,0.5,0}
\definecolor{darkblue}{rgb}{0,0,0.5}
\definecolor{lapislazuli}{rgb}{0.15, 0.38, 0.61}
\definecolor{venetianred}{rgb}{0.78, 0.03, 0.08}
\definecolor{bleudefrance}{rgb}{0.19, 0.55, 0.91}
\definecolor{dogwoodrose}{rgb}{0.84, 0.09, 0.41}
\usepackage[linktocpage,colorlinks]{hyperref}
\hypersetup{colorlinks=true, citecolor=darkgreen, linkcolor=blue,urlcolor = darkblue}
\makeatletter
\def\btt#1{\texttt{\@backslashchar#1}}
\DeclareRobustCommand\bblash{\btt{\@backslashchar}} \makeatother

\RequirePackage{fix-cm}

\begin{document}
\title{Geodesics and Bending of Light around a BTZ Black Hole Surrounded by Quintessential Matter
}
\author{Shubham Kala $^{a}$}\email{shubhamkala871@gmail.com}
\author{ Hemwati Nandan $^{a,b}$}\email{hnandan@associates.iucaa.in}
\author{Prateek Sharma $^{a}$}\email{prteeksh@gmail.com}
\author{Maye Elmardi $^{c}$}\email{maye.elmardi@gmail.com}
\affiliation{$^{a}$Department of Physics, Gurukula Kangri (Deemed to be University), Haridwar 249 404, Uttarakhand, India}
\affiliation{$^{b}$Center for Space Research, North-West University, Mahikeng 2745, South Africa}
\affiliation{$^{c}$ Center for Space Research, North-West University, Potchefstroom 2520, South Africa}


\begin{abstract}
\noindent Various observations from cosmic microwave background radiation (CMBR), type Ia supernova and baryon acoustic oscillations (BAO) are strongly suggestive of an accelerated expansion of the universe which can be explained by the presence of mysterious energy known as dark energy. The quintessential matter coupled with gravity minimally is considered one of the possible candidates to represents the presence of such dark energy in our universe. In view of this scenario, we study the geodesic of massless particles as well as massive particles around a (2+1) dimensional BTZ black hole (BH) spacetime surrounded by the quintessence. The effect of parameters involved in the deflection of light by such a BH spacetime is investigated in detail. The results obtained are then compared with a usual non-rotating BTZ BH spacetime.

\keywords{	BTZ black hole; geodesic; bending of light; quintessential matter}
\end{abstract}
\maketitle
\section{Introduction}

\noindent In general relativity (GR), a black hole (BH) solution in $ (2+1) $ dimensions with a negative cosmological constant is obtained by Bañados-Teitelboim-Zanelli (BTZ) popularly known as BTZ BH spacetime \cite{banados1992black}. This solution is different from usual $ (3+1) $ dimensional solutions in GR such as Schwarzschild BH and Kerr BH in various aspects such as it is asymptotically Anti-de-Sitter (AdS) rather than of asymptotically flat also it has no curvature singularity \cite{Carlip:1995qv}. Though, the observation support the presence of a positive cosmological constant which accelerates the universal expansion, the geometry of BTZ BH in lower dimensional gravity (i.e. the three dimensional gravity \cite{Ida:2000jh}) is quite interesting especially in presence of quintessential matter mimicking the presence of dark energy for an accelerated expansion of our universe \cite{riess1998observational,zehavi1999evidence}. Furthermore, in lower dimensional GR, the Newtonian limit does insignificant and no propagating degree of freedom appear. However, the appearance of event horizon and Hawking radiation remarked this BH solution as an authentic solution and it has also played an crucial role in the development of string theory \cite{Sfetsos:1997xs,Hyun:1997jv}. Furthermore, one of the main applications to study the properties of a BTZ BH
spacetime is to investigate the structure of higher dimensional BH spacetimes in string/M theory since their geometry at horizon are similar to the BTZ BH solutions \cite{hyun1999statistical}. Moreover, the AdS/CFT correspondence can answer such gravity dual for a particular type of quantum field theory (QFT) \cite{maldacena1999large}.The study of gravitational
lensing in view of AdS/CFT correspondence has been performed in greater detail for
Schwarzschild- AdS4 BH Spacetime which demonstrates that the holographic images
gravitationally lensed by the BH can be constructed from the
response function with an argument that the radius of Einstein rings depends
on the total energy of the QFT thermal state \cite{hashimoto2020imaging}.\\

Type-Ia supernovae measurements \cite{kippenhahn1990stellar,leibundgut2001cosmological,sullivan2006rates,maoz2014observational} and other observations such as BAO, CMBR suggest an accelerated expansion of the universe. In the frame work of GR, such an accelerated phase of universe is expected to be driven by an exotic kind of matter with negative pressure called dark energy \cite{Peebles:2002gy,Wang:2016lxa}. Dark energy acts as a repulsive gravitational force so that usually it is modelled as an exotic fluid. So far, variety of dark energy models with dynamical scalars field such as quintessence \cite{Kiselev:2002dx,AzregAinou:2012hy,Azreg-Ainou:2014lua,Zhou:2007xp,Zhou:2007xp}, k-essence \cite{Yang:2009zzl,Guo:2004fq}, quintom \cite{xia2006features} and phantom dark energy \cite{Kunz:2006wc,bouali2019cosmological} have been proposed as an alternative models to the cosmological constants. Quintessence is a spatially in-homogeneous component with negative pressure and is a possible candidate for dark energy, which is characterized by an equation of state \cite{Steinhardt:2003st}. The solution
for a spherically symmetric spacetime geometry surrounded by a quintessence matter is first studied by Kiselev \cite{Kiselev:2002dx}. The various other properties of BHs surrounded by quintessence have been studied extensively in diverse contexts in recent times  \cite{AzregAinou:2012hy,Azreg-Ainou:2014lua,Thomas:2012zzc,Azreg-Ainou:2014lua,Fernando:2014wma,Fernando:2014rsa,Azreg-Ainou:2017obt}. \\ 

The study of geodesic has played a significant role to obtain a better understanding the physical properties of a BH \cite{cruz1994geodesic}. In general, the solutions of geodesic equations are obtained analytically wherever possible while in the absence of analytical solutions, these equations are solved numerically as well. Despite the complexity of analytic methods, due to its higher accuracy, there are lot of studies that have been done by using the Weierstrassian elliptic function and Kleinian sigma function \cite{whittaker2020course,eichler1982zeros,abramowitz1970handbook} in near past. The analytical solutions for many well-known spacetimes i.e. Schwarzschild \cite{hagihara1930theory}, four-dimensional Schwarzschild de-Sitter \cite{Hackmann:2008zz}, higher–dimensional Schwarzschild, Schwarzschild–(anti) de -Sitter, Reissner–Nordstrom and Reissner–Nordstrom–(anti)-de-Sitter \cite{hackmann2008analytic}, Kerr \cite{kerr1963gravitational}, Kerr–de-Sitter \cite{Hackmann:2010zz}, Reissner Nordstrom–(anti)-de-Sitter surrounded by regular and exotic matter field \cite{Chatterjee:2019rym}, three dimensional rotating BH \cite{Kazempour:2017gho}, (2+1) dimensional charged BTZ BH \cite{Soroushfar:2015dfz} and BH in f(R) gravity \cite{Soroushfar:2015wqa} have been studied previously.\\

Further, it is also well known that the deflection of light (i.e. bending of light) is one of useful tool to search for the dark and massive objects in our universe. Recently, several attempts have been made to study the gravitational lensing (GL) within weak as well as strong deflection limit \cite{Iyer:2009wa,Bozza:2009yw,Virbhadra:1999nm} around such objects. The geodesic structure of the quantum-corrected Schwarzschild BH surrounded by quintessence has also been studied recently (see \cite{Nozari:2020tks}). It is important to analyze the deflection of light in lower dimensional BH spacetime surrounded by quintessential matter. In this study, we take into account a (2+1)-dimensional BTZ BH surrounded by quintessential matter to study the structures of correspondent null and time-like geodesics respectively along with the bending of light around this particular BH spacetime. We also compare the results of present study with the usual BTZ BH in GR.\\

The overview of present paper is as follows. In section 2, we put forward the metric and than obtain the corresponding geodesic equations and effective potential for null as well as timelike geodesics respectively. Section 3 includes the analytical solutions of geodesic equations for massless and massive particles along with the classification of possible orbits in dependence upon the energy parameter and the angular momentum parameter of the test particles. In section 4, the exact expression of bending angle of light is derived and the effect of quintessential matter in bending angle is analyzed accordingly. Finally, we conclude the results obtained in section 5.
\section{BTZ BH Surrounded by Quintessential Matter, Geodesic Equations and Effective Potential}

\noindent The line element ansatz for a $(2 + 1)$-dimensional planar BH may be written as \cite{banados1999three}
\begin{equation}
	ds^2 = -A(r)dt^2 + A(r)^{-1}dr^2 + r^2d{\phi}^2, \label{m1}
\end{equation}
where r stands for the radial coordinate such that $0 < r < \infty$ and $\phi$ is a planar coordinate -$\infty < \phi < \infty$. 
The function A(r) in Eq. \ref{m1}  is determined by the Einstein’s field equations,
\begin{equation}
	R_{ab}- \frac{1}{2}g_{ab}R-\frac{1}{l^2}g_{ab} = 8{\pi}T_{ab}, \label{e2}
\end{equation}

here $l$ standing for the AdS radius which is related to the cosmological constant $(\Lambda)$ as $l^2 = \frac{-1}{\Lambda}$, and $T_{\mu \nu}$ is the energy-momentum tensor for the quintessential matter. Following \cite{Chen:2012mva,Kiselev:2002dx} such a tensor can be caste in terms of the quintessence energy density $(\rho_{q})$ and the state parameter of the quintessence, $(w_{q})$ as below, 

\begin{equation}
	T^{t}_{\phantom{t}t} = T^{r}_{\phantom{r}r} = -\rho_{q} , \hspace{0.5cm} T^{\phi}_{\phantom{\phi}\phi} = (2w_{q}+1)\rho_{q}.  \label{e3}
\end{equation}

Solving Einstein’s field equations Eq. \ref{e2} with the line element ansatz Eq. \ref{m1} and the energy-momentum given by Eq. \ref{e3} , one has the metric for a BTZ BH spacetime in terms of Schwarzschild coordinates as below \cite{deOliveira:2018weu},

\begin{equation}
	ds^2 = \frac{-r^2}{l^2}f(r)dt^2 + \frac{l^2}{r^2}f(r)^{-1} dr^2 + r^2 d{\phi^2} ,\label{metric1}
\end{equation}
where, 
$f(r) = 1-(\frac{r_+}{r})^{\sigma} $ , $ {\sigma} = 2(1+w_q)$ and       $r_+ = (Ml^2)^\frac{1}{\sigma}$.
Here, $r_{+}$ specify the event horizon with $M$ being the BH mass, $w_{q}$ represents the quintessential parameter for equation of state (EOS) which lies between -1 to 0. The line element in Eq. \ref{m1} can also be obtained, with the proper choice of charges, as a particular solution from a generic description of quintessential BHs given by Kiselev \cite{Kiselev:2002dx}. In particular, for $w_{q} = 0$, one can recover the metric of the usual BTZ BH \cite{banados1992black} in GR. One can obtain geodesic equations using Lagrangian equation corresponding to a given spacetime is below,

\begin{multline}
	\mathcal{L} = \frac{1}{2} \Sigma^{3}_{\mu,\nu=0}, g_{\mu\nu}\frac{dx^\mu}{d\lambda}\frac{dx^\nu}{d\lambda} = \frac{1}{2}\epsilon \\ \hspace{-1cm} = \frac{1}{2} \left[\frac{-r^2}{l^2}f(r)\left(\frac{dt}{d\lambda}\right)^2 + \frac{l^2}{r^2}f(r)^{-1}\left(\frac{dr}{d\lambda}\right)^2 + r^2\left(\frac{d\phi}{d\lambda}\right)^2   \right], \label{L}
\end{multline}

where the value of $\epsilon$ assigned 1 and 0 for massive and massless particles respectively and $\lambda$ is an affine parameter. One can obtain the constants of motion assisted by Euler–Lagrange equation,

\begin{equation}
	P_t = \frac{{\delta}L}{\delta\dot{t}} = \frac{-r^2}{l^2}f(r)\dot{t} = -E, P_\phi = \frac{{\delta}L}{\delta\dot{\phi}} = r^2\dot{\phi} = L, \label{com}
\end{equation}

where $E$ and $L$ represent the Energy and angular momentum of test particle respectively. With the use of Eq. \ref{L} and Eq. \ref{com}, one can obtain the geodesic equations as following,

\begin{equation}
	\left(\frac{dr}{d\lambda}\right)^2 = E^2 + \frac{r^2}{l^2}f(r)\epsilon - \frac{L^2}{l^2}f(r), \label{ger}
\end{equation}

\begin{equation}
	\left(\frac{dr}{d\phi}\right)^2 = \epsilon\frac{f(r)}{l^2L^2}r^6 + \left(\frac{E^2}{L^2}- \frac{f(r)}{l^2}\right)r^4, \label{gephi}
\end{equation}

\begin{equation}
	\left(\frac{dr}{dt}\right)^2 = \frac{{\left\lbrace f(r)\right\rbrace}^3 }{E^2l^6}{\epsilon}r^6 + \left[ \frac{{\left\lbrace f(r)\right\rbrace}^2}{l^4} - \frac{L^2{\left\lbrace f(r)\right\rbrace}^3}{E^2l^6}\right]r^4. \label{get}
\end{equation}

\begin{figure}[h]
	\centering
	\includegraphics[width=6cm,height=5cm]{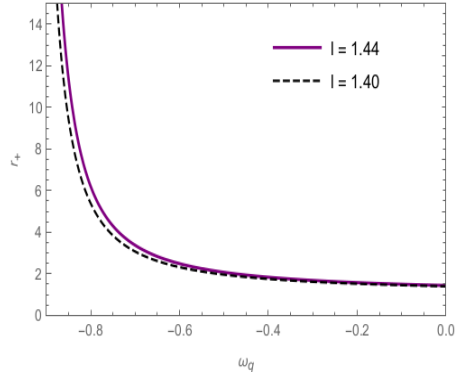}
	
	\caption{Variation of event horizon with radial quintessence parameter for different values of cosmological constant for $ M = 1 $. } \label{hor}
\end{figure}

The effective potential then given by,

\begin{equation}
	V_{eff} = \frac{L^2}{l^2}f(r) - \frac{r^2}{l^2}{\epsilon}f(r). 
\end{equation}

For null geodesics $(\epsilon=0)$ and timelike geodesics $(\epsilon=1)$, the effective potential reads as,

\begin{equation}
	V_{eff(null)} = \frac{L^2}{l^2}f(r), \hspace{0.8cm} V_{eff(timelike)} = \frac{L^2}{l^2}f(r) - \frac{r^2}{l^2}f(r),
\end{equation}

above expressions represent the effective potential of null and timelike geodesics respectively for BTZ BH surrounded by the quintessence. In order to study, the photon orbits i.e, motion of photons around BH, we analyse the variation of effective potential in view of radial distance ($ r $) and angular momentum ($ L $) for a massless particle. The spacetime in Eq. \ref{metric1} is basically characterized by $w_q$ and $\Lambda$.\\

 Here we consider $w_q=-1/2$ to study the geodesics and lensing phenomena because for other values of $w_q$, the equations of motion are cumbersome and one can not have the analytical solutions of these equations. The given limit of EOS parameter $w_q=-1/2$ represents a quintessence model of dark energy (since for quintessence it adhere to $ -1 < w_q < -1/3) $. We have therefore considered this particular value to obtain analytic solutions for the study of geodesic equations and an exact solution for deflection angle of light around the BH accordingly. However, other values are only considered to study the variation of effective potential since it is difficult to solve the concerned governing equations analytically. For $w_q=-1/2$, the metric function $f(r) = \left(1-\frac{Ml^2}{r}\right)$ and the Eq. \ref{gephi} therefore leads to

\begin{equation}
	\left(\frac{dr}{d\phi}\right)^2 = \left(\frac{1}{l^2L^2}\right){\epsilon}r^6 - \frac{M}{L^2}r^5 + \left(\frac{E^2}{l^2}-\frac{1}{l^2}\right)r^4-Mr^3 = R(r). \label{e12}
\end{equation}

Here, $R(r)$ is a polynomial of function $r$. For convenience, the dimensionless parameters are defined as

\begin{equation}
	\tilde{r}=\frac{r}{M} , \tilde{l}=\frac{l}{M} , \tilde{L}=\frac{M^2}{L^2}, \label{e13}
\end{equation}

With Eq. \ref{e13}, the Eq. \ref{e12} is then rewritten as,

\begin{equation}
	\left(\frac{d\tilde{r}}{d\phi}\right)^2 = \left(\frac{\tilde{L}}{\tilde{l}^2}\right){\epsilon}\tilde{r}^6 - \tilde{L}M^2{\tilde{r}^5} + \left(\frac{E^2}{\tilde{l}}-\frac{1}{\tilde{l}^2}\right)\tilde{r}^4 + M^2\tilde{r}^3 = \frac{R(\tilde{r})}{M^2}. \label{e14}
\end{equation}

\subsection  {Null geodesics}
\begin{figure}[h]
	\subfigure[]{\includegraphics[width=6cm,height=5cm]{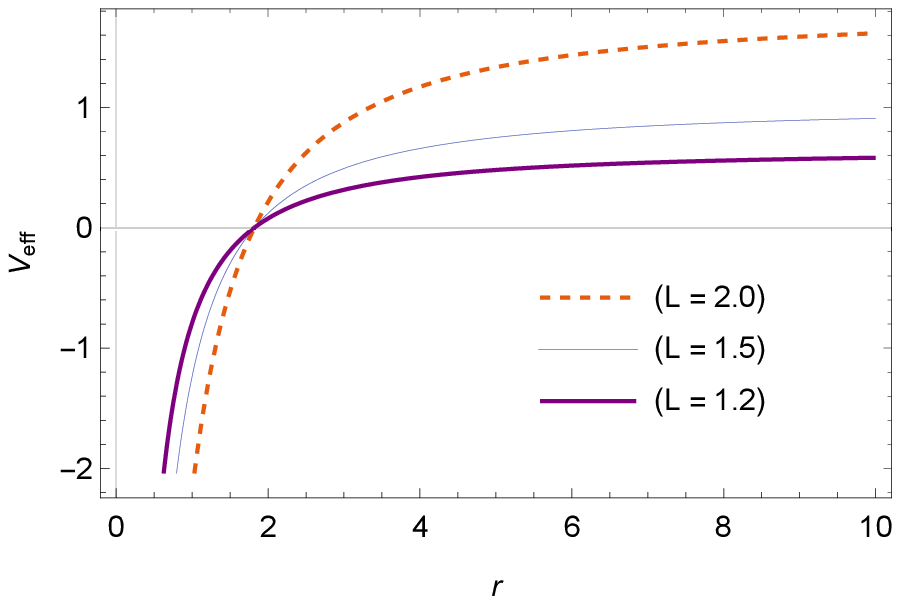}} \label{ep1}
	\centering
	\subfigure[]{\includegraphics[width=6cm,height=5cm]{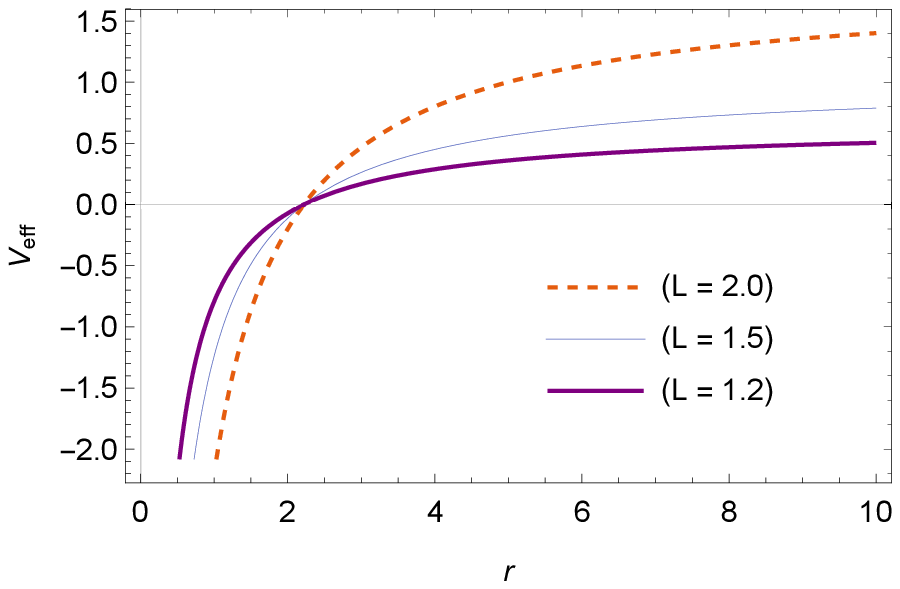}} \label{ep2}
	\caption{Effective potential for various values of $L$ when $ M = 1 $ with (a) $w_{q}=-1/3$ (left panel) and (b) $w_{q}=-1/2$ (right panel).} \label{epd}
\end{figure}
\begin{figure}[h]
	\subfigure[]{\includegraphics[width=6cm,height=5cm]{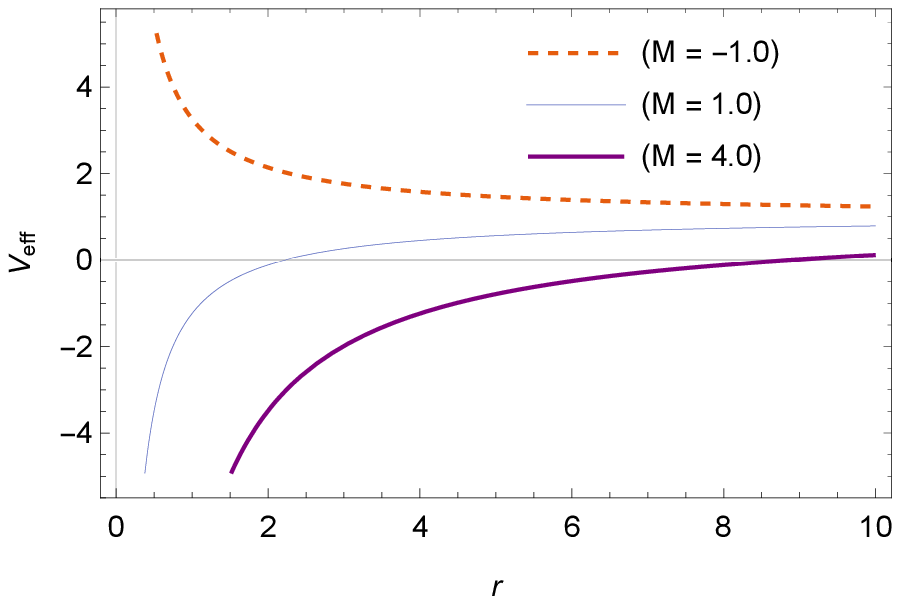}} \label{ep3}
	\centering
	\subfigure[]{\includegraphics[width=6cm,height=5cm]{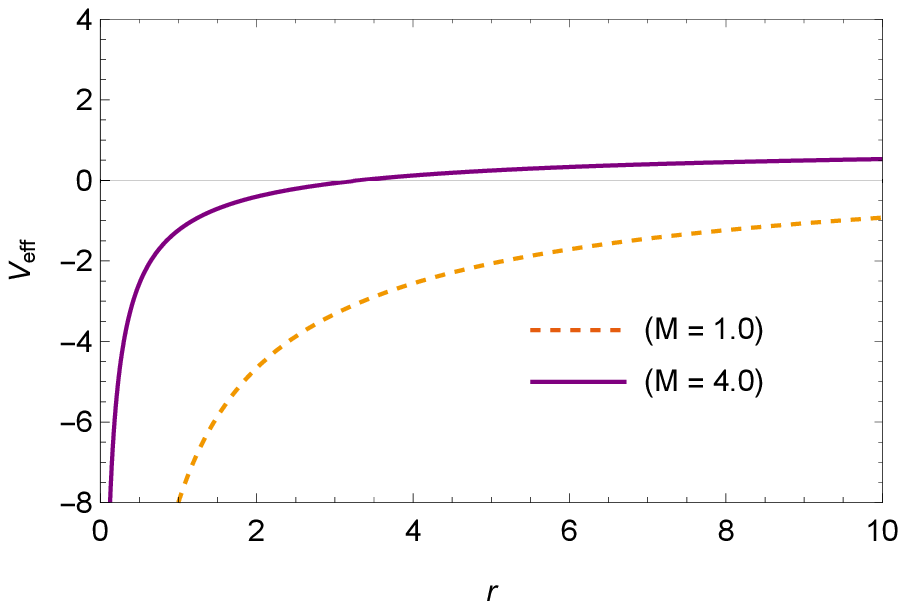}} \label{ep4}
	\caption{Effective potential for various values of $M$ when $ L = 1.5 $ with (a) $w_{q}=-1/2$ (left panel) and (b) $w_{q}=-2/3$ (right panel). } \label{epm}
\end{figure}
\begin{figure}[h]
	\begin{center}
		\subfigure[]{\includegraphics[width=6cm,height=5cm]{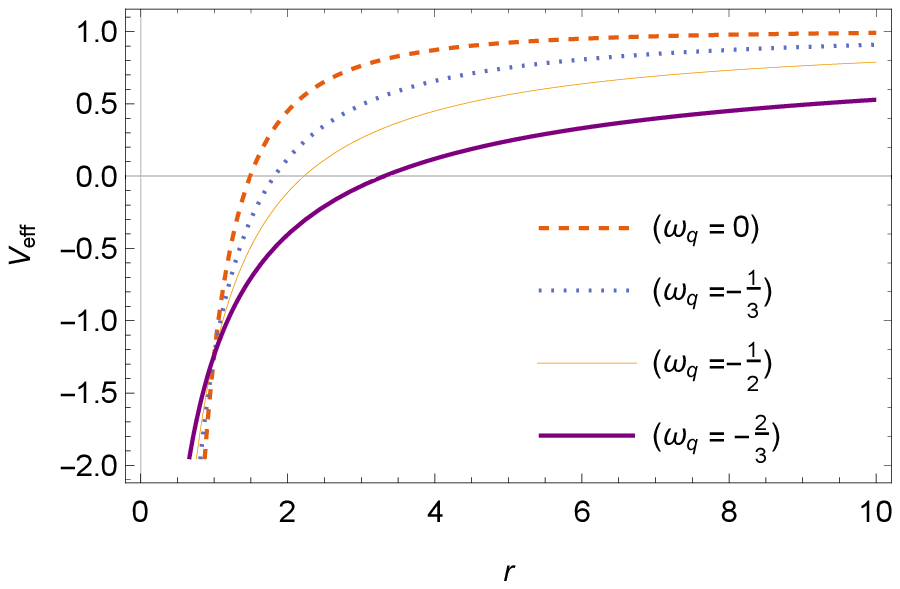}} \label{ep5}
		\subfigure[]{\includegraphics[width=6cm,height=5cm]{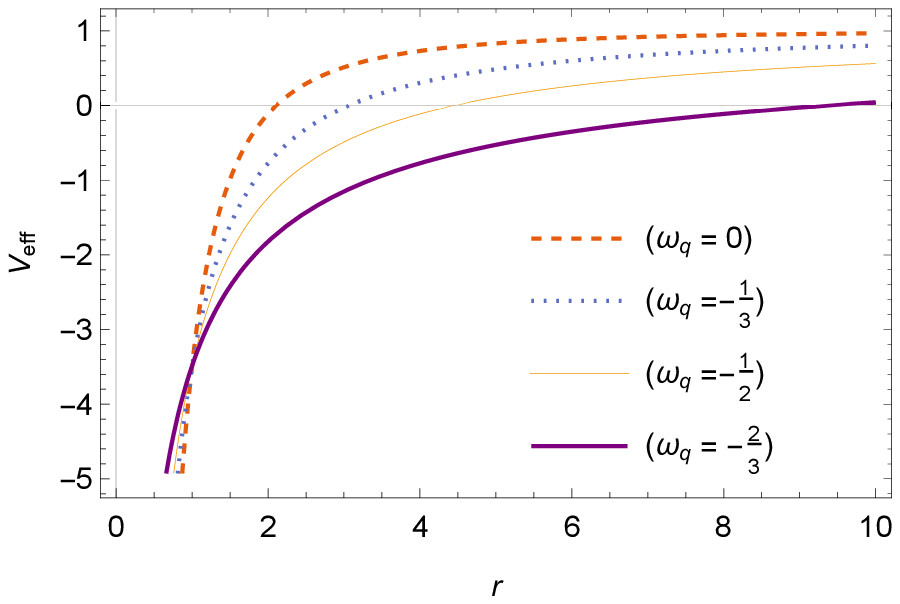}} \label{ep5b}
	\end{center}
	
	\caption{Effective potential for various values of  $\omega_{q}$ with (a) $ M=1 $ (left panel) and (b) $ M=2 $ (right panel).} \label{epw} 
\end{figure}
\noindent The variation of effective potential with radial distance for different values of angular momentum of photons is presented in  \figurename{ \ref{epd}} for fixed $M$ and for different value of $M$ in \figurename{ \ref{epm}} respectively. The effective potential for the different values of angular momentum for photons, have neither maxima nor minima and also, there is no turning point. The only possible orbit may therefore be terminating orbits. However, on increasing the value of quintessence parameter, the zero for effective potential shift towards higher values. It is observed that the photon can escape from BH when $ M  > 0 $ and  $ E \geq L $. Moreover, for $ M < 0 $ (i.e.  AdS geometry) the photons can escape to infinity and there is a lower bound for radial coordinate. It is also observed that as $ w_{q} $ decreases the effective potential for $ M > 0 $ shifts towards negative region which in turn suggests that an inwardly directed particles always plunge into the singularity. It can be easily noticed from  \figurename{ \ref{epw}} that the value of effective potential decreases with the decline in quintessence parameter $w_q$. 

\subsection  {Timelike geodesics}
\begin{figure}[h]
	\subfigure[]{\includegraphics[width=6cm,height=5cm]{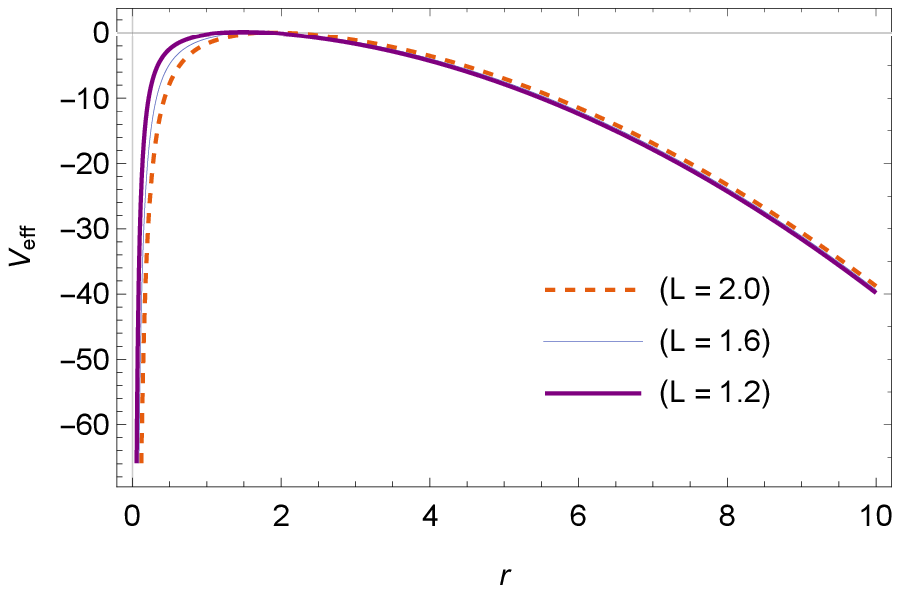}} \label{ep6}
	\centering
	\subfigure[]{\includegraphics[width=6cm,height=5cm]{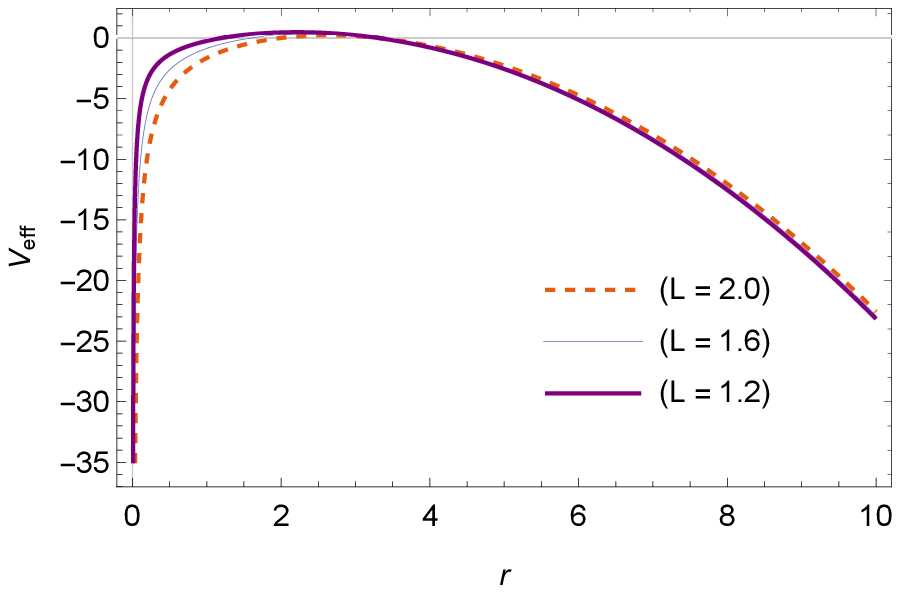}} \label{ep7}
	\caption{Effective potential for various values of $L$ when $ M = 1 $ with (a) $w_{q}=-1/2$ (left panel) and  (b) $w_{q}=-2/3$ (right panel).}\label{epd1}
\end{figure}
\begin{figure}[h]
	\subfigure[]{\includegraphics[width=6cm,height=5cm]{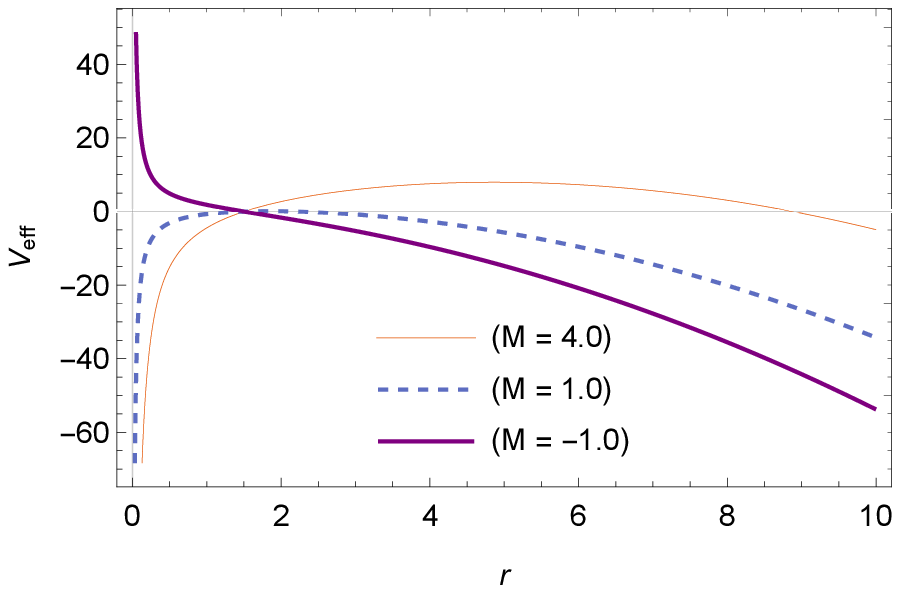}} \label{ep8}
	\subfigure[]{\includegraphics[width=6cm,height=5cm]{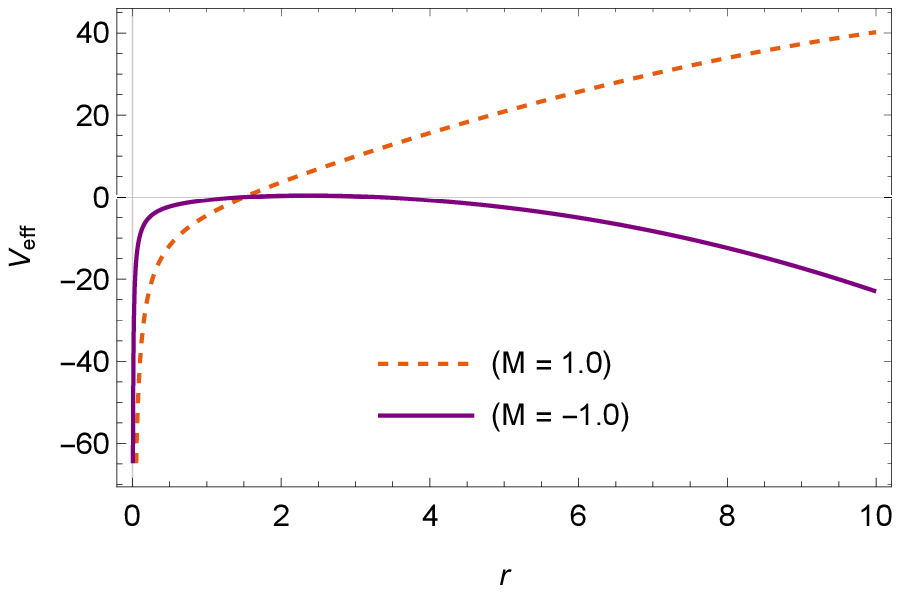}} \label{ep9}
	\caption{Effective potential for various values of $M$ when $ L = 1.5 $ with (a) $w_{q}=-1/2$ (left panel) and (b) $w_{q}=-2/3$ (right panel). }\label{epm1}
\end{figure}
\begin{figure}[h]
	\begin{center}
		\subfigure[]{\includegraphics[width=6cm,height=5cm]{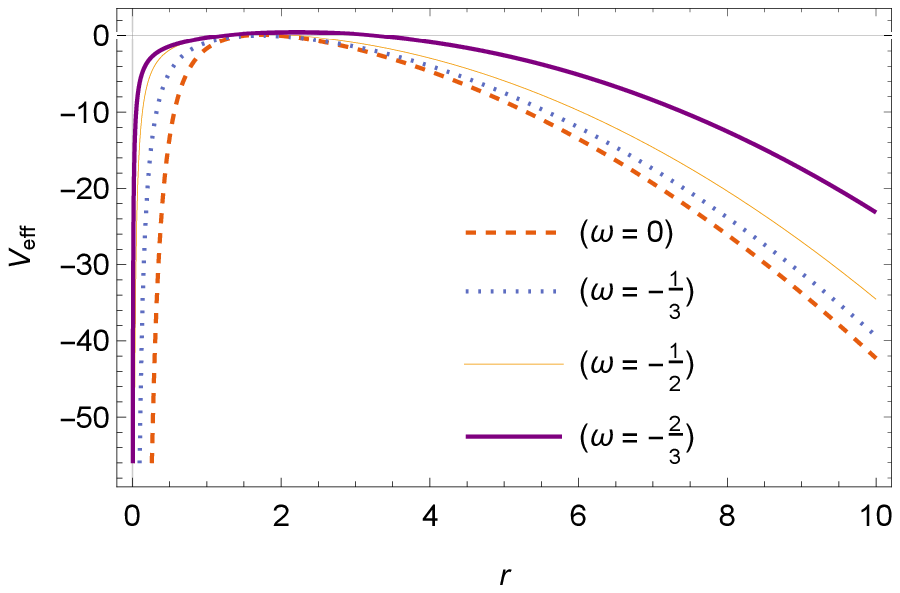}} \label{ep10}
		\subfigure[]{\includegraphics[width=6cm,height=5cm]{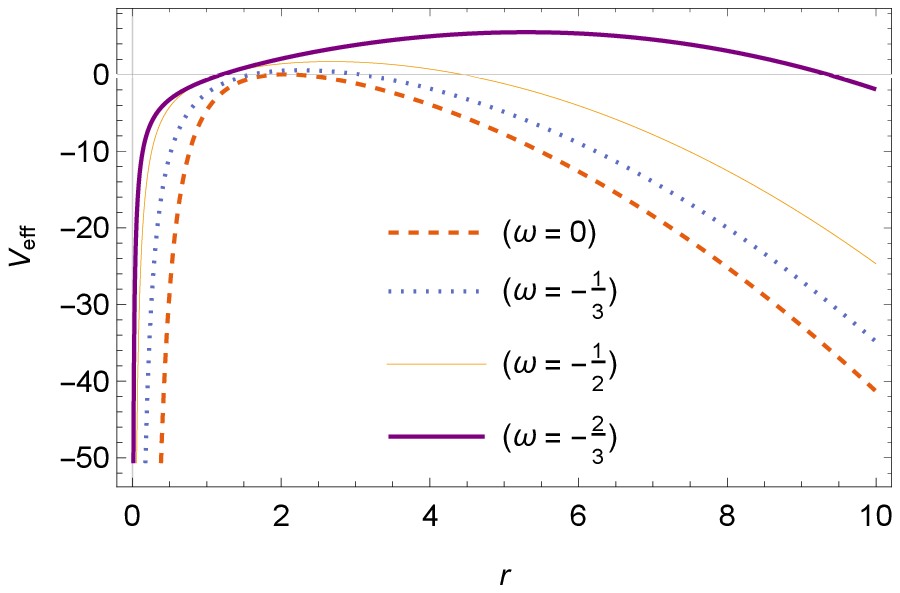}} \label{ep10b}
	\end{center}
	\caption{Effective potential for various values of  $\omega_{q}$ with (a) $ M =1 $ (left panel) and  (b) $ M = 2 $ (right panel).} \label{epw2} 
\end{figure}
\noindent The variation of effective potential with radial distance for different values of angular momentum of test particles is presented in \figurename{ \ref{epd1}} and for different value of $M$ in \figurename{ \ref{epm1}} respectively. The effective potential for the various values of angular momentum for test particles, have only maxima for certain values of $r$. Also, there is one turning point for the test particles and the possible orbits may therefore be terminating orbits and bound orbits. The behaviour of effective potential for different values of $M$ is opposite to the massless particles, i.e. for Ads geometry, the test particle can not escape to infinity and beyond the horizon for $M > 0$, the effective potential shifts towards the positive region. It can be easily observed from \figurename{ \ref{epw2}} that the value of effective potential decreases with the decrease in $w_q$.

\subsection {Comparison to other cases of parameter $w_q$}

For $w_q=-1/3$, the metric function is reads as $f(r)= 1- \left\lbrace \frac{(Ml^2)^\frac{3}{4}}{r}\right\rbrace^\frac{4}{3}$,  so we have

\begin{equation}
	\left(\frac{dr}{d\phi}\right)^2 = \frac{\epsilon(1-Ml^2)}{l^2L^2}(r)^\frac{22}{3}-\frac{(1-Ml^2)}{l^2}(r)^\frac{8}{3} + \frac{E^2}{L^2}. \label{e15}
\end{equation}

For $w_q=-2/3$, the metric function is $f(r)= 1- \left\lbrace \frac{(Ml^2)^\frac{3}{2}}{r}\right\rbrace^\frac{2}{3}$, so we have

\begin{equation}
	\left(\frac{dr}{d\phi}\right)^2 = \frac{\epsilon(1-Ml^2)}{l^2L^2}(r)^\frac{16}{3}-\frac{(1-Ml^2)}{l^2}(r)^\frac{10}{3} + \frac{E^2}{L^2}. \label{e16}
\end{equation}

However for $w_q=-1$, the metric function vanishes such that 

\begin{equation}
	\left(\frac{dr}{d\phi}\right)^2 = \frac{E^2}{L^2}r^4.
\end{equation}

Eq. \ref{e15} and Eq. \ref{e16} contain some terms with fractional powers of r. To  the best of our knowledge, it is not easy to solve such equation analytically and may be solved numerically \cite{Hartmann:2010rr}. 

\subsection  {Possible regions for Geodesic motion}
Eq. \ref{e14} asserts that $R(\tilde{r}) > 0$ is a necessary condition for the existence of a geodesic motion. Therefore, the real positive zeros of  $R(\tilde{r})$ are in fact the extremal values of the geodesic motion and figure out the type of geodesic. Since $\tilde{r} = 0$ is a zero of this polynomial for all values of the parameters, one can neglect it. So the  Eq. \ref{e14} changes to a polynomial of third degree as follows,

\begin{equation}
	R^*(\tilde{r}) = \left(\frac{\tilde{L}}{\tilde{l}^2}\right)\tilde{r}^3 - \tilde{L}M^2{\tilde{r}^2} + \left(E^2\tilde{L}-\frac{1}{\tilde{l}^2}\right)\tilde{r} + M^2.
\end{equation}

One can analyse the nature of possible orbits with the use of analytical solutions, which depend on the parameters of massive test particles or light rays (i.e. $\epsilon, E^2, l$ and $L$) as discussed in the next section. Considering a set of parameters $\epsilon, E^2, l$ and $L$, the polynomial $R^*(r)$ has a several number of positive and real zeros which can be identified by Descarte's rule. The number of zeros can change only if two zeros of  $R^*(r)$  merge to one and the parameters $E^2$ and $L$ are varied accordingly. Finding solution $R^*(\tilde{r})=0$ and $\frac{dR^*(\tilde{r})}{d\tilde{r}}=0$, provide us $E^2$ and $\tilde{L}$. For massive test particles $(\epsilon=1)$, we have

\begin{equation}
	\hspace{1.5cm}  \tilde{L} = \frac{M^2\tilde{l}^2}{(2-M^2\tilde{l}^2)\tilde{r}^2}, \hspace{2cm} E^2 =\frac{4\tilde{r}^2-2M^2l^2\tilde{r}^2-6M^2\tilde{l}^2\tilde{r}+3M^4\tilde{l}^4\tilde{r}+M^4\tilde{l}^4}{2M^2\tilde{l}^4}.
\end{equation}

However, for massless particles $(\epsilon=0)$,

\begin{equation}
	\tilde{L} =\frac{1}{E^2}\left(\frac{1}{\tilde{l}^2}-\frac{M^2}{\tilde{l}}\right).
\end{equation}

\begin{figure} [h]
	\begin{center}
		\subfigure[]{\includegraphics[width=6cm,height=5cm]{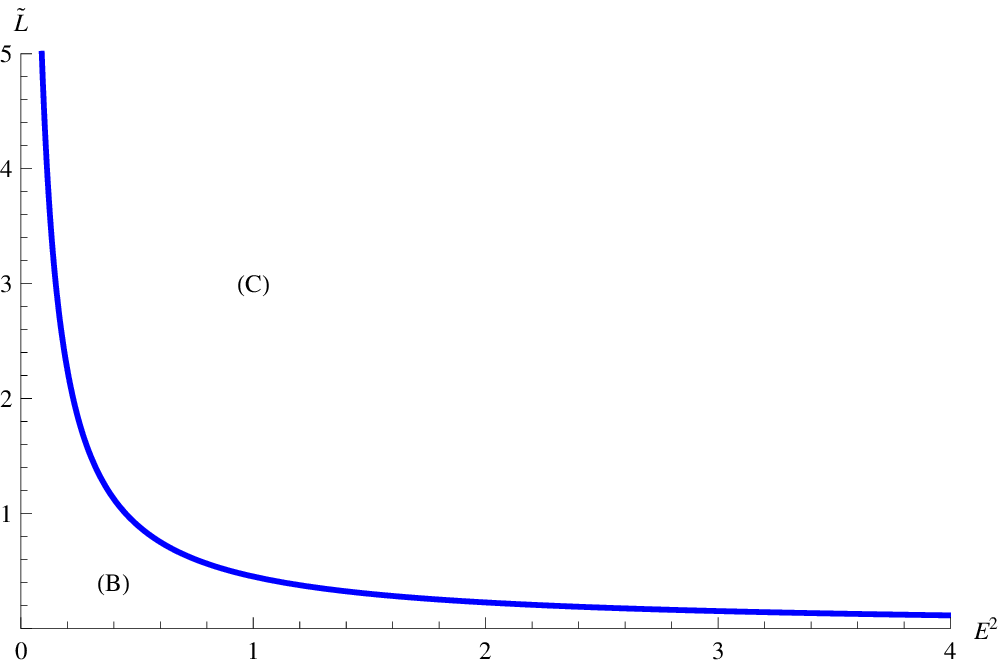}} 
		\subfigure[]{\includegraphics[width=6cm,height=5cm]{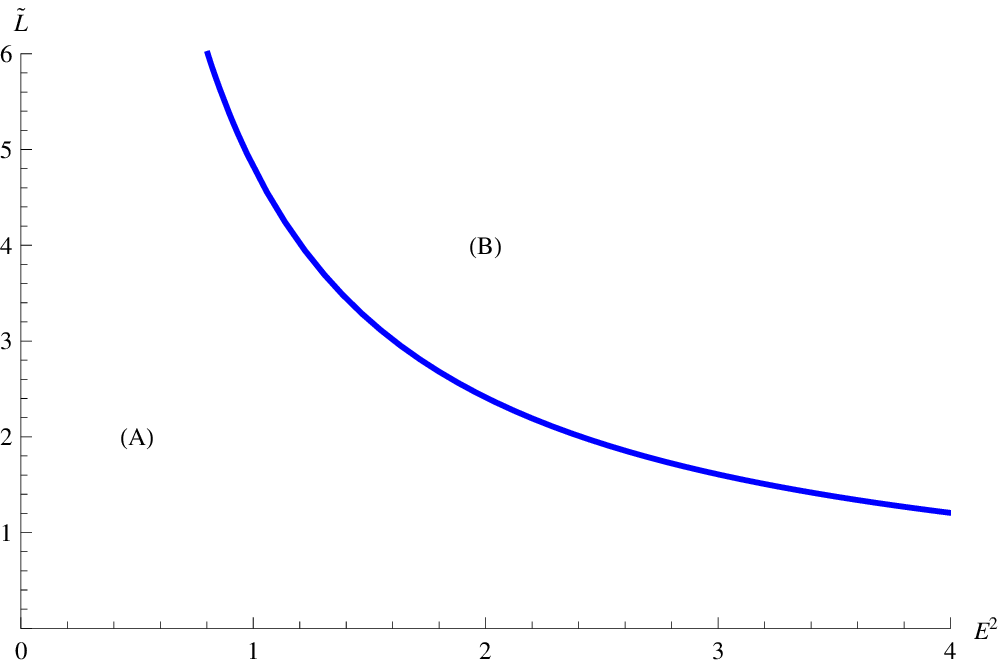}}
	\end{center}
	\caption{Different types of geodesic motion wherein the geodesic motion is possible: (a) for light rays $\epsilon = 0$ (b) for massive particles $\epsilon = 1$. The numbers of positive real zeros in these regions are: A=2, B = 1, C = 0.} \label{prog}
\end{figure}

The regions of different types of geodesic motion are classified accordingly (see \figurename{ \ref{prog}}).

\section{Analytical Solution of Geodesic Equations}

In the present section, we study analytical solutions of geodesic equations. Using a new parameter $u=\frac{1}{\tilde{r}}$, we simplify  Eq. \ref{e14} as below,
\begin{equation}
	\left(\frac{du}{d\phi}\right)^2 = \frac{\tilde{L}\epsilon}{\tilde{l}^2u^2}-\frac{\tilde{L}M^2\epsilon}{u} + \left(E^2\tilde{L}-\frac{1}{\tilde{l}^2}\right)+M^2u, \label{e21}
\end{equation}
which can be solved for both the null and timelike geodesics as discussed below.

\begin{table*}[ht] 
	\begin{center}
		\caption{Types of orbits of light and timelike particles in the vicinity of BTZ BH surrounded by quintessential matter. The range of the orbits depicted here by horizontal lines however the turning points are shown by thick dots. The double vertical line represent the position of horizon.} \label{table}
		\begin{tabular}{|c|c|c|c|}
			\hline
			Region	& Zeros  & Range of $\tilde{r}$  & Orbits \\
			
			\hline
			A	& 0 &  \begin{tikzpicture}[
				dot/.style = {circle, fill=black,inner sep=0pt, minimum size=4pt},
				every label/.append style = {inner sep=0pt, rotate around={25:(-0.8,1.5)}},
				thick      
				]
				\draw ( 0.0,-0.2) -- + (0.0,0.4) node[below] {};
				\draw[thick] (0.1,0.001) -- node[below=1mm] {} + (0.19,0);
				\draw[thick] (0.3,0.001) -- node[below=1mm] {} + (0.19,0);
				\draw ( 0.6,-0.2) -- + (0,0.4) node[below] {};
				\draw ( 0.65,-0.2) -- + (0,0.4) node[below] {};
				%
				\draw[thick] (0.65,0.001) -- node[below=1mm] {} + (4,0);
				\draw[thick] (4,0.001) -- node[below=1mm] {} + (4,0);
				\foreach \p in {0.25, 0.5}
				{
					\node[dot,label={}] at (10*\p,0) {};
				}
			\end{tikzpicture} & TEOs,EOs\\
			
			\hline
			B	& 2 &  \begin{tikzpicture}[
				dot/.style = {circle, fill=black,inner sep=0pt, minimum size=4pt},
				every label/.append style = {inner sep=0pt, rotate around={25:(-0.5,1.5)}},
				thick      
				]
				\draw ( 0.0,-0.2) -- + (0.0,0.4) node[below] {};
				\draw[thick] (0.1,0.001) -- node[below=1mm] {} + (0.19,0);
				\draw[thick] (0.3,0.001) -- node[below=1mm] {} + (0.19,0);
				\draw ( 0.6,-0.2) -- + (0,0.4) node[below] {};
				\draw ( 0.65,-0.2) -- + (0,0.4) node[below] {};
				%
				\draw[thick] (0.65,0.001) -- node[below=1mm] {} + (4,0);
				\draw[thick] (4,0.001) -- node[below=1mm] {} + (4,0);
				\foreach \p in {0.35}
				{
					\node[dot,label={}] at (10*\p,0) {};
				}
			\end{tikzpicture} & BOs   \\
			
			\hline
			C	& 1 &  \begin{tikzpicture}[
				dot/.style = {circle, fill=black,inner sep=4pt, minimum size=4pt},
				every label/.append style = {inner sep=0pt, rotate around={25:(0.5,0.5)}},
				thick      
				]
				\draw ( 0.0,-0.1) -- + (0,-0.4) node[below] {};
				\draw[thick] (0.1,-0.25) -- node[below=2mm] {} + (0.19,0);
				\draw[thick] (0.3,-0.28) -- node[below=2mm] {} + (0.19,0);
				\draw ( 0.6,-0.1) -- + (0,-0.4) node[below] {};
				\draw ( 0.65,-0.1) -- + (0,-0.4) node[below] {};
				%
				\draw[thick] (0.65,-0.3) -- node[below=2mm] {} + (4,0);
				\draw[thick] (4,-0.3) -- node[below=2mm] {} + (4,0);
				%
			\end{tikzpicture} & TBOs \\
			\hline
		\end{tabular}
		
	\end{center}
\end{table*}

\subsection{Null geodesics}
For $\epsilon=0$,  Eq. \ref{e21} reduces to,

\begin{equation}
	\left(\frac{du}{d\phi}\right)^2 = M^2u + \left(E^2\tilde{L}-\frac{1}{\tilde{l}^2}\right), \label{e22}
\end{equation}

which is a polynomial of less than order three and hence the analytic solution of equation is not possible to obtain in terms of Weistrass fuction \cite{eichler1982zeros}. The analytic solution of above equation is however given as

\begin{equation}
	r({\phi}) = \frac{4}{M \phi} - M^2\left(\frac{1}{E^2 \tilde{L}}- l^2\right).
\end{equation}

Using this solution, one can identify the null geodesics for each regions of different types of orbits as discussed in the section of classification of the orbits..

\subsection{Timelike geodesics}
For $\epsilon=1$,  Eq. \ref{e21} reduces to,

\begin{equation}
	\left(\frac{du}{d\varphi}\right)^2 = \frac{\tilde{L}}{\tilde{l}^2u^2}-\frac{\tilde{L}M^2}{u} + \left(E^2\tilde{L}-\frac{1}{\tilde{l}^2}\right)+M^2u. \label{e24}
\end{equation}

For convince, above equation can also be rewritten as,

\begin{equation}
	\left(u\frac{du}{d\varphi}\right)^2 = M^2u^3 + \left(E^2\tilde{L}-\frac{1}{\tilde{l}^2}\right)u^2-{\tilde{L}M^2u} +\frac{\tilde{L}}{\tilde{l}^2} =P_3(u),
\end{equation}

the nature of above equation represent an elliptic type. With use of substitution $u=\frac{1}{a_3}\left(4y-\frac{a_2}{3}\right)
=\frac{1}{M^2}\left[4y-\frac{(E^2-1/\tilde{l}^2)}{3}\right]$, transforms
Eq. \ref{e24} into Weierstrass form as follow,

\begin{equation}
	\left(\frac{dy}{d\varphi}\right)^2 = 4y^3-{\zeta}y-{\eta} = P_3(y) \label{e26}
\end{equation} 

wherein,

\begin{multline}
	\zeta = \frac{{a_2}^2}{12} - \frac{a_1a_3}{4} = \frac{3\tilde{L}M^4+\left(E^2\tilde{L}-\frac{1}{\tilde{l}^2}\right)}{12}, \\ 
	\eta =\frac{{a_1}{a_2}{a_3}}{48} - \frac{a_0{a_3}^2}{16} - \frac{{a_2}^3}{216}  \\=\frac{-\tilde{L}M^4\left(E^2\tilde{L}-\frac{1}{\tilde{l}^2}\right)}{48}-\frac{\tilde{L}M^4}{\tilde{l}16} - \frac{\left(E^2\tilde{L}-\frac{1}{\tilde{l}^2}\right)}{216},
\end{multline}

\noindent are widely known as  Weierstrass constants. The Eq. \ref{e26} is of an elliptic type and can be solved by using the Weierstrass function \cite{Hackmann:2008zz,Soroushfar:2015wqa} as indicated below,

\begin{equation}
	y(\varphi) = \wp(\varphi-{\varphi}_{in};\zeta,\eta),
\end{equation}

$ where  $${\varphi}_{in}={\varphi}_{0} + \int_{y_{0}}^{\infty}\frac{dy}{\sqrt{4y^3-{\zeta}y-{\eta}}}$ and $y_0= \frac{1}{4}\left(\frac{a_3}{\tilde{r_0}}+\frac{a_2}{3}\right) =\frac{1}{4}\left\lbrace\frac{M^2}{\tilde{r_0}}+\frac{1}{3}\left(E^2\tilde{L}-\frac{1}{\tilde{l}^2}\right)\right\rbrace $ and it depends only the initial values $\phi_0$ and $\tilde{r_0}$. The analytical solution of Eq. \ref{e14} in form of Weierstrass function then reads as,

\begin{multline}
	\tilde{r_0}(\varphi) = \frac{a_3}{4\wp(\varphi-{\varphi}_{in};\zeta,\eta)} \\ = \frac{M^2}{4\wp(\varphi-{\varphi}_{in};\zeta,\eta)-\frac{1}{3}\left(E^2\tilde{L}-\frac{1}{\tilde{l}^2}\right)}.
\end{multline}

Using this solution, one can identify the regions for different types of orbits as discussed below.

\subsection{Classification of the Orbits}

With the use of analytical solution obtained in previous section, the effective potential in accordance with the zeroes of underlying polynomials, one can visualize the structure of the possible types of orbits in different regions (see \tablename{ \ref{table}}) as specified below,\\
\textbf{Region A}: No real zeroes appeared, Escape Orbits (EO) and Terminating Escape Orbits (TEO) are possible in this region.\\
\textbf{Region B}: Two zeroes come into existence, both the zeros are positive and they are located at outside the horizon. Therefore, we may have Bound Orbits (BO) in this region.\\
\textbf{Region C}: Only zero appear, we may have Terminating Bound Orbits (TBO) only such that the particles moving from any finite coordinate distance to $r = 0$.\\

\noindent The number of real zeros assigned with the polynomial provide us the significant description to understand the motion of light rays as well as massive test particles. In the case if polynomial does not acquire any zeros, then the  light rays and particles would come from infinity and straight away towards to the singularity at origin. This type of orbit is called as a terminating orbit. Moreover, one real positive zero represent the particle coming from a finite coordinate distance and ends at origin, which so called as a terminating bound orbit. If the polynomial possessed two real zeroes, we can have two cases: $(i)$ the light rays keep moving forward on an escape orbit and do not cross origin. $(ii)$ there is a bound orbit and it would cross the origin several times. The brief description of possible orbit types is categorized in \tablename{ \ref{table}}.\\

\begin{figure}[h]
	
	\subfigure[]{\includegraphics[width=6cm,height=5cm]{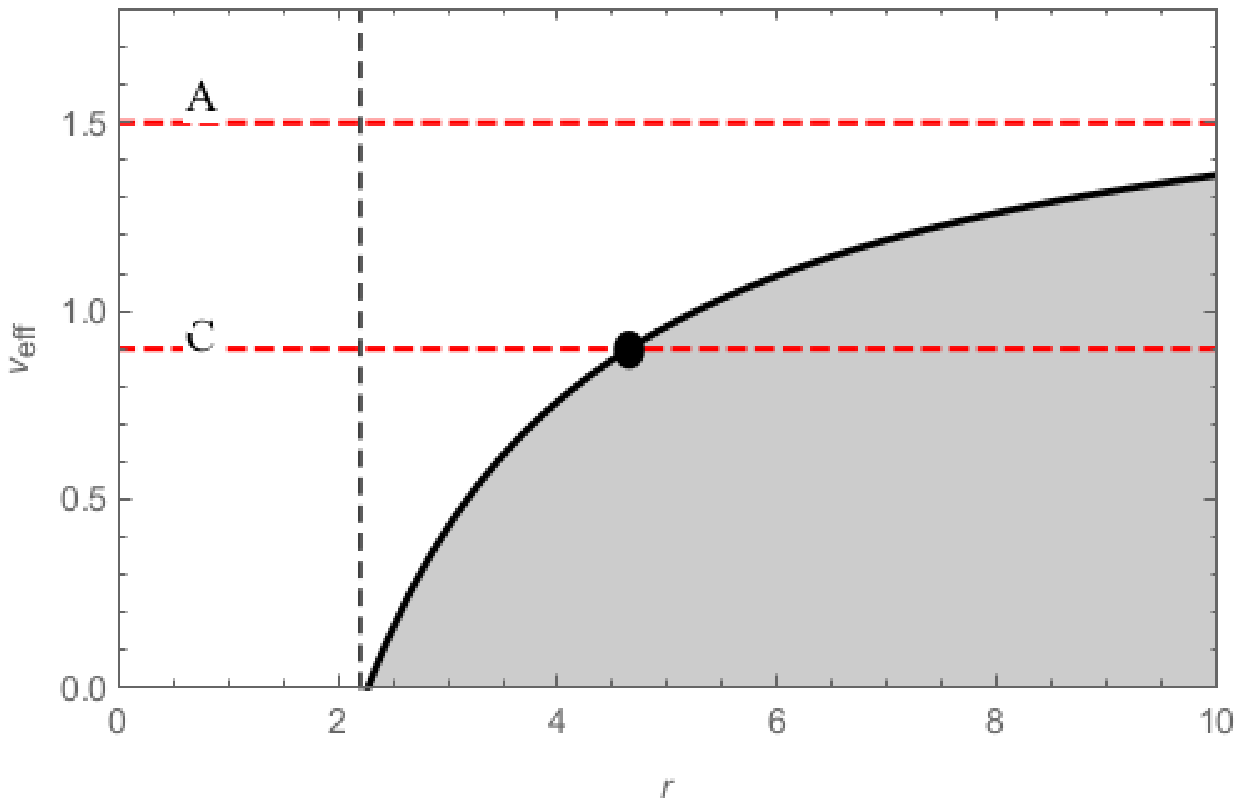}}
	\subfigure[]{\includegraphics[width=6cm,height=5cm]{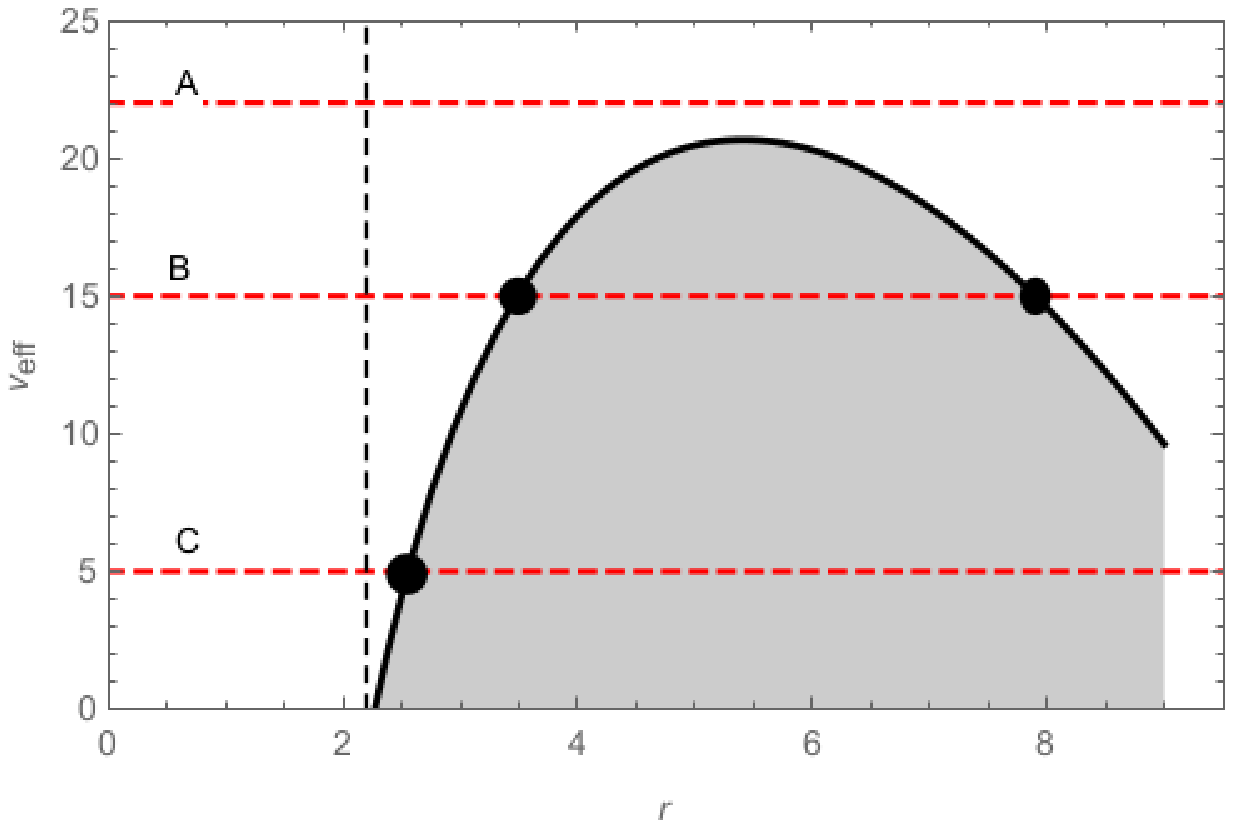}}
	\centering
	\caption{ $(a)$ Effective potential for light rays with parameters $\epsilon=0$,$M=1$,$L=2$, and $l=1.49$. 
		$(b)$ Effective potential for massive particles with parameters $\epsilon=1$,$M=1$,$L=2$, and $l=1.49$.}\label{no}
	
\end{figure}

\begin{figure}[h]
	\subfigure[]{\includegraphics[width=6cm,height=5cm]{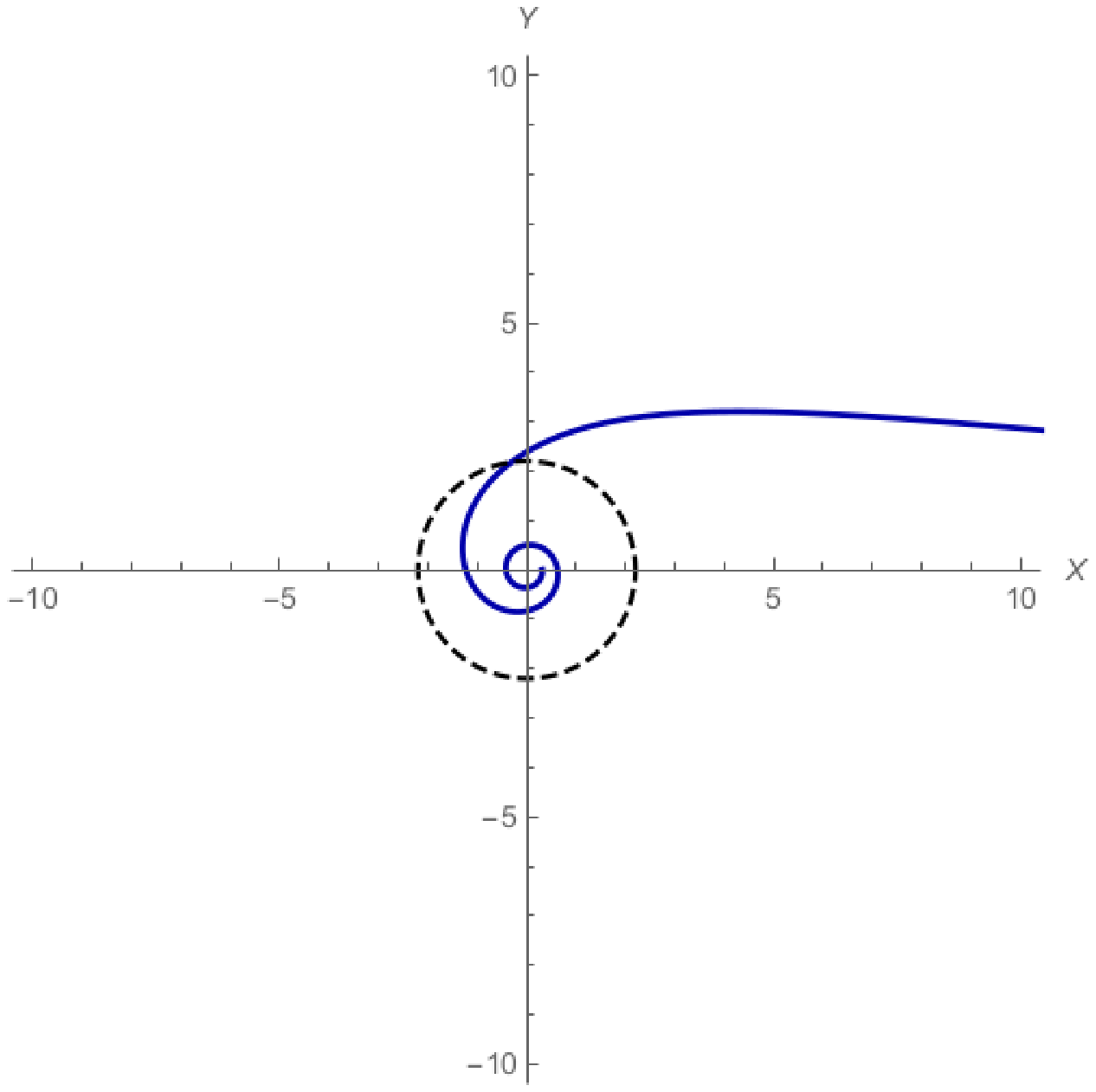}}
	\subfigure[]{\includegraphics[width=6cm,height=5cm]{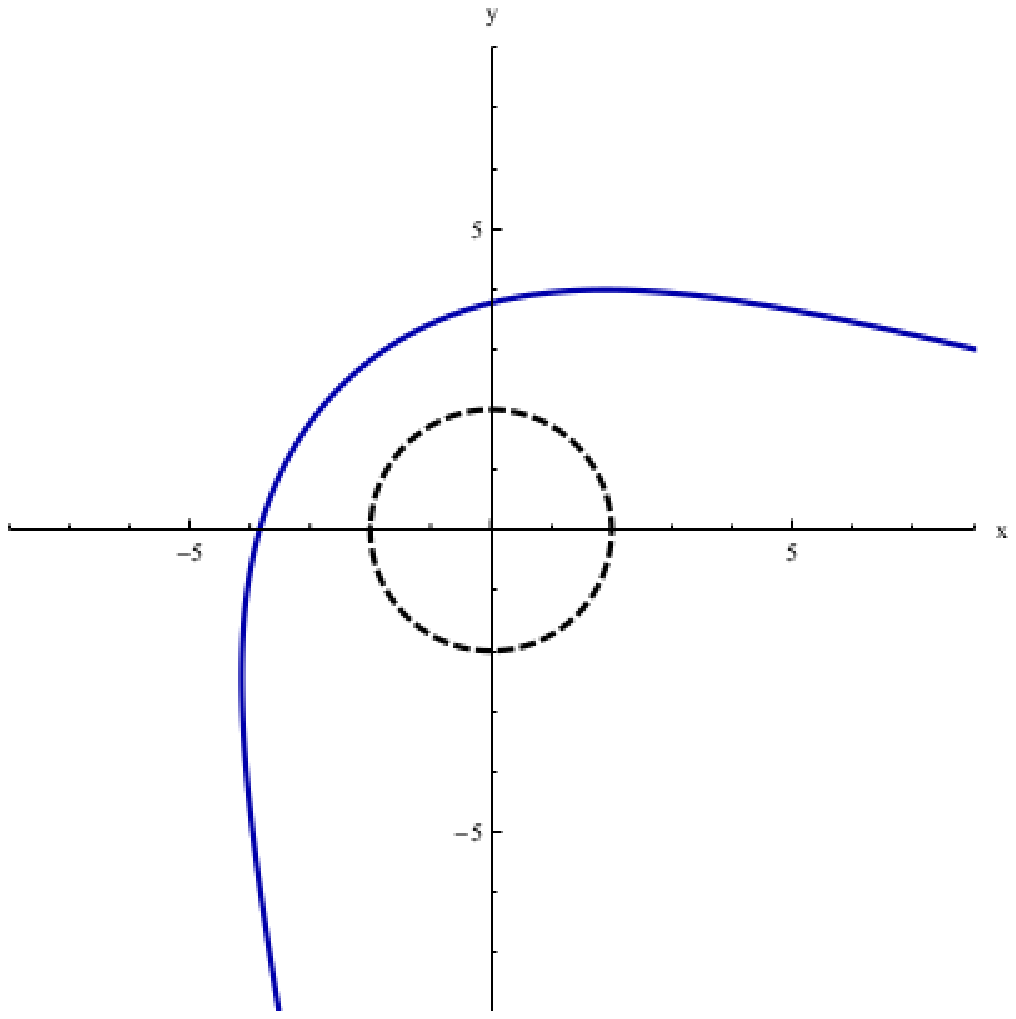}}
	\subfigure[]{\includegraphics[width=6cm,height=5cm]{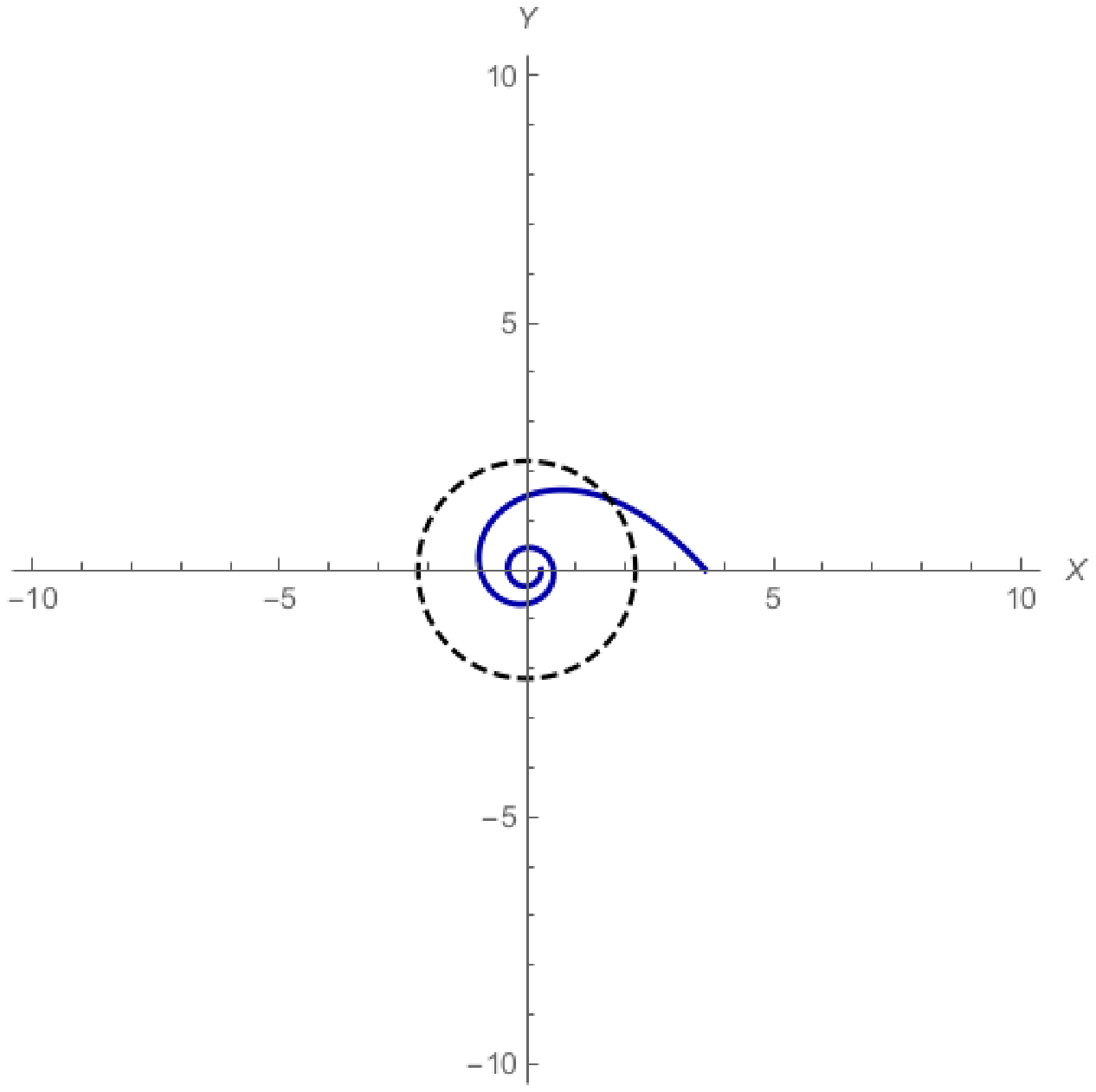}}
	\centering
	\caption{Orbits of BTZ BH surrounded by quintessence  for light rays: (a) For region A; $E^2=1.1$,$L=2$, $M=1$ and $l= 1.49$: TEO, (b) For region A; $E^2=0.5$,$L=0.9$, $M=1$ and $l= 1.49$: EO and (C) For region C; $E^2=2.5$,$L=2$, $M=1$ and $l= 1.49$: TBO.} \label{nuo}
\end{figure}

\begin{figure}[h]
	\subfigure[]{\includegraphics[width=6cm,height=5cm]{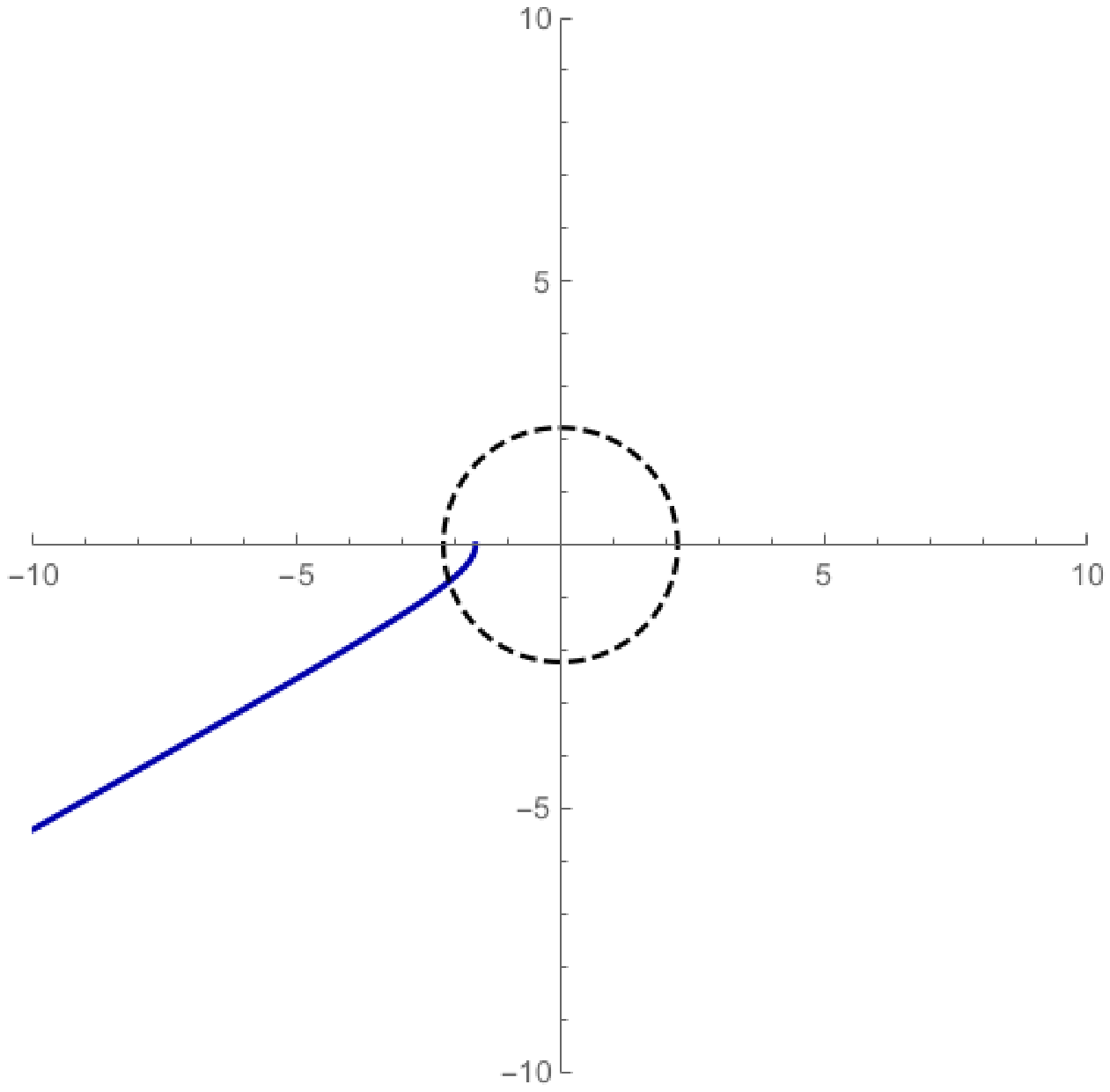}}
	\subfigure[]{\includegraphics[width=6cm,height=5cm]{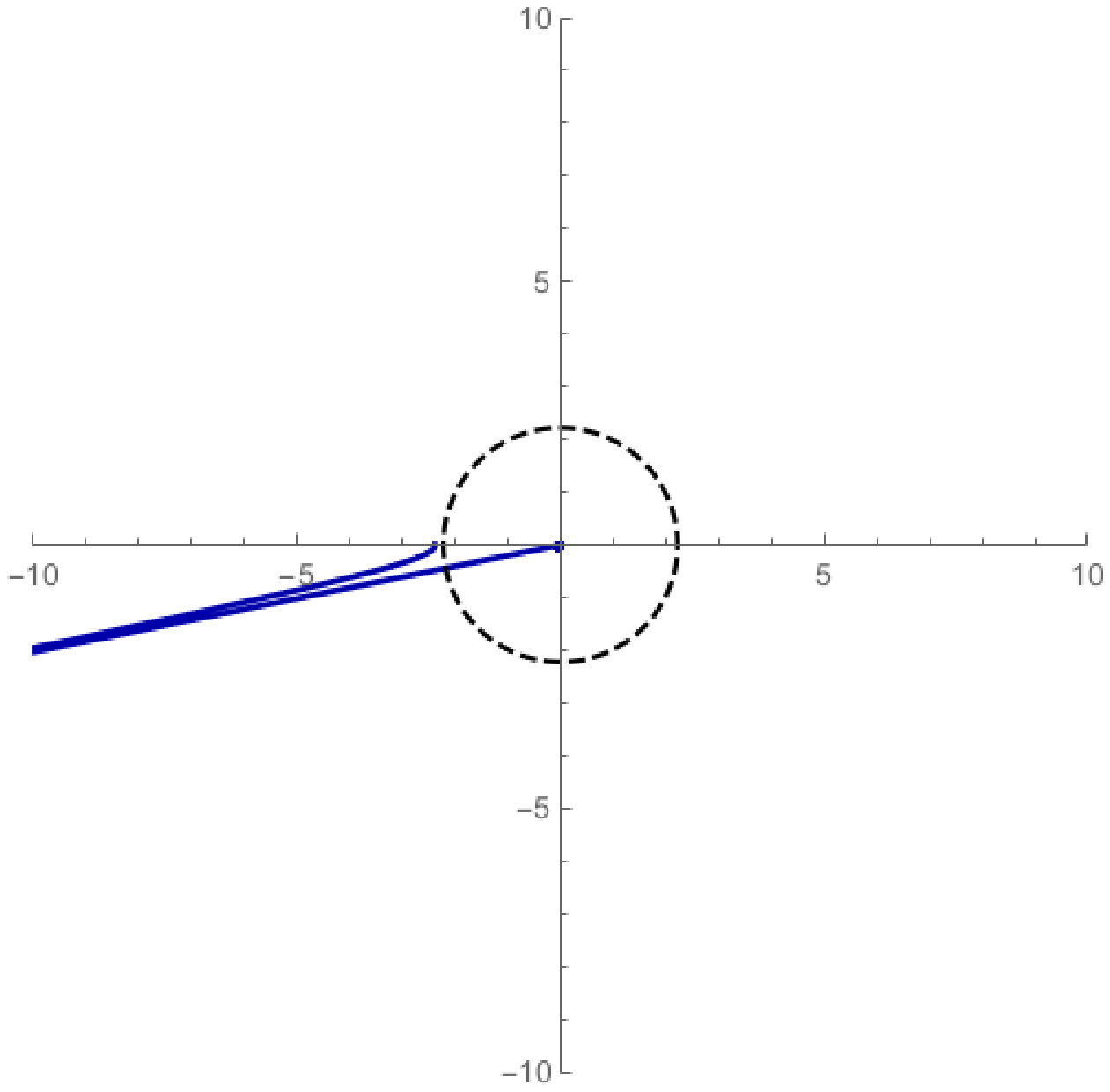}}
	\centering
	\subfigure[]{\includegraphics[width=6cm,height=5cm]{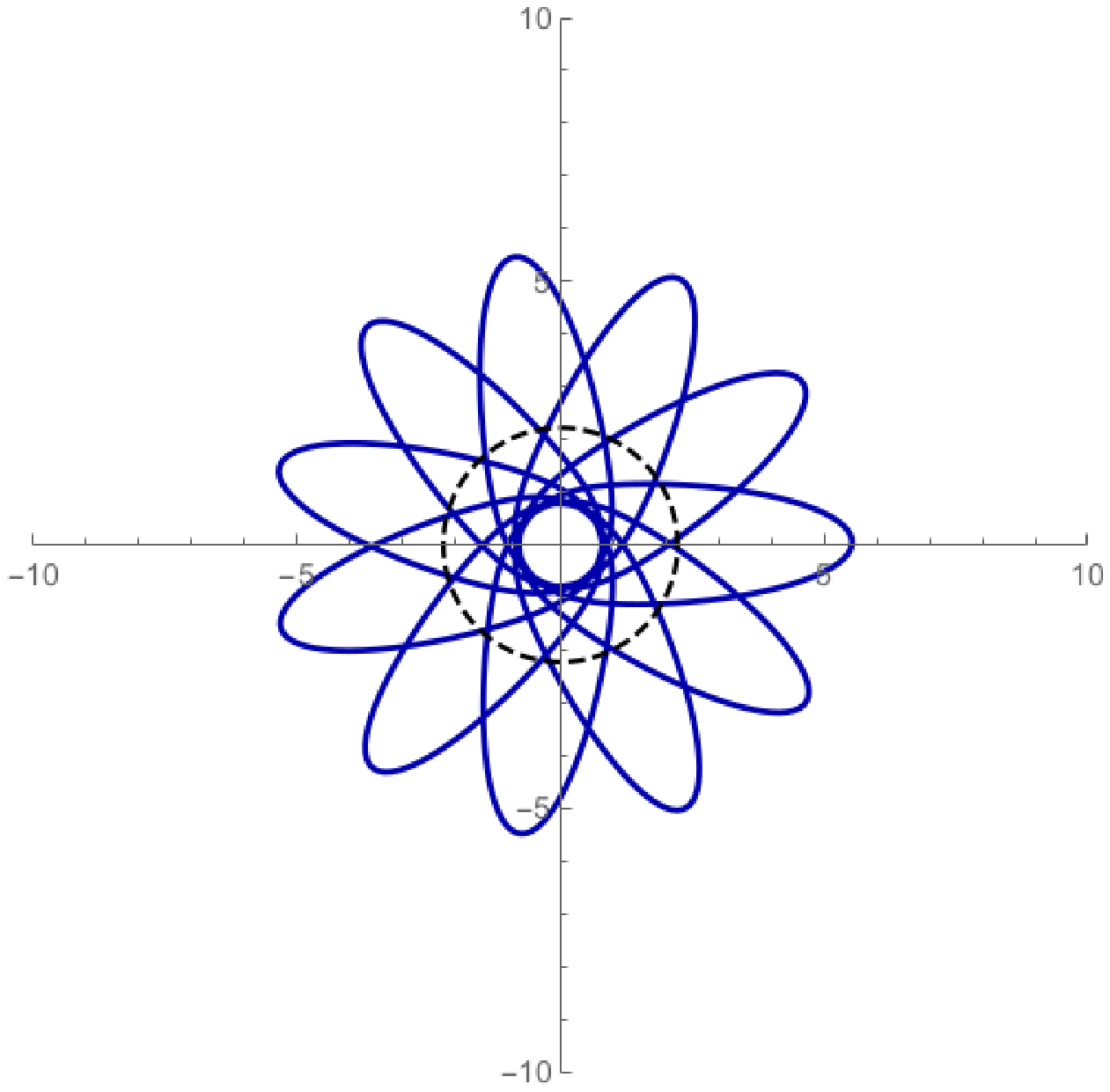}}
	\caption{Orbits of BTZ BH surrounded by quintessence  for massive particles: (a) For region A; $E^2=0.5$,$L=2$, $M=1$ and $l= 1.49$: TEO, (b) For region C; $E^2=0.5$,$L=0.9$, $M=1$ and $l= 1.49$: TBO and (c) For region B; $E^2=2$,$L=1.33$, $M=1$ and $l= 1.49$: BO. } \label{tlo}
\end{figure}

\noindent The effective potential for light rays and test particles and the possible orbits are shown in \figurename{ \ref{no}} where the vertical black dashed line represents the horizon of BH. The red dashed horizontal lines  represent the energies of regions A, B and C  and the black dots mark which are the turning points of the orbits indicate the zeros of polynomial R. Further, the horizon radii has represented by the vertical double line. The shaded area is a forbidden zone, where there is no geodesic motion possible. The analytical results obtained indeed allow us to identify some specific sets of orbits for massless and massive test particles respectively in the BTZ BH surrounded by quintessential matter. In concordance with the parameters  $l$, $M$, $L$ and $E$ terminating escape orbit, escape orbit and terminating bound orbits for massless particles can be visualized in \figurename{ \ref{nuo}}. However, the terminating escape orbit and bound orbits for massive test particles are shown in \figurename{ \ref{tlo}}.

\section{The Exact Bending Angle}

One can consider a beam of photons approaching the BH which is coming from the asymptotic region. From asymptotic region the appearing light ray reaches upto an observer with $r_{0}$ as the distance of closest approach . At the closest approach distance $r_{0}$, the light rays deflected by its strong attraction to reach to a far observer in another point (in an asymptotically flat region). Now take an account the definition of bending of light, we know that, the change in $\phi$ and the deflection angle $\delta$ is simply related by a difference of $\pi$. The angle of deflection, which is responsible for the apparent location of that remote
point for a far observer, is given as \cite{Keeton:2005jd,Uniyal:2018ngj,Sharma:2019qxd},

\begin{equation}
	\delta = - \pi + 2\int_{r_{0}}^{\infty} \left| \frac{d\phi}{dr} \right| dr. \label{e30}
\end{equation}

By using Eq. \ref{e22} and Eq. \ref{e30}, the expression of bending angle for BTZ BH surrounded by quintessence for EOS parameter $w_q=-1/2$, can be written in terms of variable $u$ as follows,

\begin{equation}
	\delta_{w_{q}} = - \pi + 2\int_{0}^{u_{0}} \left[\sqrt{\dfrac{1}{Mu} + (b^2-l^2)}\right] du,  \label{alpha}
\end{equation} 

here, $ u_{0} = \dfrac{(L^2/l^2-E^2)}{M L^2}. $ 

Above integral is solved to obtain an exact expression for bending angle of light, which can be expressed as,

\begin{multline}
	\delta_{w_{q}} = - \pi + \sqrt{b^2 - l^2 + 1/(M u_0)} u_0 \\ + 
	\frac{\log\left[{1 + 2 b^2 M u_0 - 2 l^2 M u_0} + 
		2 \sqrt{b^2 - l^2} M \sqrt{b^2 - l^2 + 1/(M u_0)} u_0\right]}{2 \sqrt{b^2 - l^2} M} ,  \label{alpha}
\end{multline}

\noindent where, $b^2=\frac{L^2}{E^2}$ and other symbols have their usual meanings. The above expression can be simplified after some arrangement of constant term i.e. $b' =\sqrt{b^2 -l^2}$ and it becomes,

\begin{multline}
	\hspace{2cm} \delta_{w_{q}} = - \pi + \sqrt{b' + \frac{1}{M u_0}} u_0 \\ + 
	\frac{\log\left[{1 + 2 M b'^2 u_0 - 2 M b' u_0} \left(1+\sqrt{1+\frac{1}{Mb'u_0}}\right)\right]}{2b'M}.
\end{multline}

\noindent By using the expansion of logarithm function in above expression, the exact expression of bending angle can be written as,

\begin{equation}
	\delta_{w_{q}} = - \pi + \left[b'+\xi\left(\sqrt{b'}+1\right)\right]u_0 - Mb'\left(b'+\xi\right)^2 u_{0}^2 , \label{ba1}
\end{equation}

where,  $\xi = \sqrt{1+\frac{1}{Mb'u_0}} $.
For $ \omega_{q} = 0 $,  the above spacetime $(1)$ reduces to the usual BTZ BH spacetime case as given below,

\begin{equation}
	ds^2 = \frac{-r^2}{l^2}f(r)dt^2 + \frac{l^2}{r^2}f(r)^{-1} dr^2 + r^2 d{\phi^2} ,\label{metric}
\end{equation}

where, 
$f(r) = \left(1-\frac{Ml^2}{r^2}\right)$.\\

The expression of bending angle for BTZ BH without quintessence then takes the following form accordingly,

\vspace{1cm}
\begin{equation}
	\delta = - \pi + 2\int_{0}^{u_{0}} \left[\sqrt{\dfrac{1}{Mu^2} + (b^2-l^2)}\right] du.  \label{alpha} 
\end{equation}
which leads to,
\vspace{2cm}

\begin{multline}
	\delta  = - \pi \\+ \frac{\sqrt{\left(b^2-l^2+\frac{1}{M{u_0}^2}\right)} {u_0}\left[\sqrt{1+Mb^2{u_0}^2-Ml^2{u_0}^2} + \log{u_0}-\log\left(1+\sqrt{1+Mb^2{u_0}^2-Ml^2{u_0}^2}\right)\right]}{\sqrt{1+Mb^2{u_0}^2-Ml^2{u_0}^2}},   \label{alphanr}
\end{multline}

\noindent The above expression can however be further simplified after rearrangement of different terms as follows,

\begin{multline}
	\hspace{2cm}	\delta  = - \pi \\ + \frac{\sqrt{\left(b'+\frac{1}{M{u_0}^2}\right)} {u_0}\left[\sqrt{1+Mb'{u_0}^2} + \log{u_0}-\log\left(1+\sqrt{1+Mb'{u_0}^2}\right)\right]}{\sqrt{1+Mb'{u_0}^2}},
\end{multline}

\noindent here, $ u_{0} = \sqrt{\frac{\left(L^2/l^2-E^2\right)}{ML^2}}$ ,   $b^2=\frac{L^2}{E^2}$ and $ b' = b^2-l^2$.\\

The exact expression of bending angle for BTZ BH in absence of quintessence i.e. usual BTZ BH thus reads as,

\begin{equation}
	\delta  = - \pi + \frac{\sqrt{b'} \xi \left[1+\log{u_0}+Mb'u_{0}^2\right]u_{0}}{2\sqrt{1 + Mb'u_{0}^2}}, \label{ba2}
\end{equation}

where $\xi = \sqrt{1+\frac{1}{Mb'u_{0}^2}}$. \\

The explicit expression for bending angle of light around BTZ BH with and without quintessence are therefore seen by Eq. \ref{ba1} and Eq. \ref{ba2} respectively.
\begin{figure}[h]
	\subfigure[]{\includegraphics[width=6cm,height=5cm]{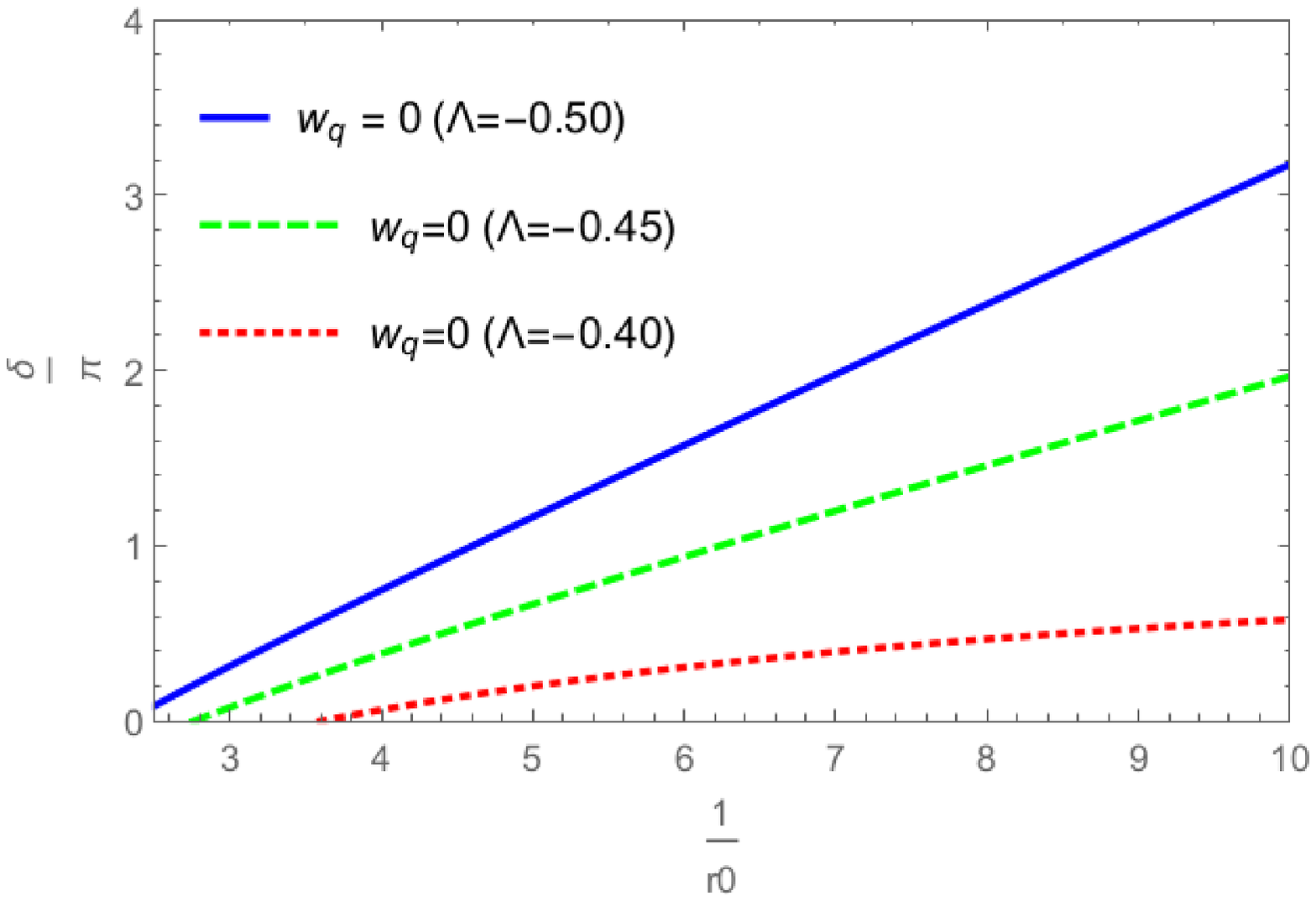}} 
	\subfigure[]{\includegraphics[width=6cm,height=5cm]{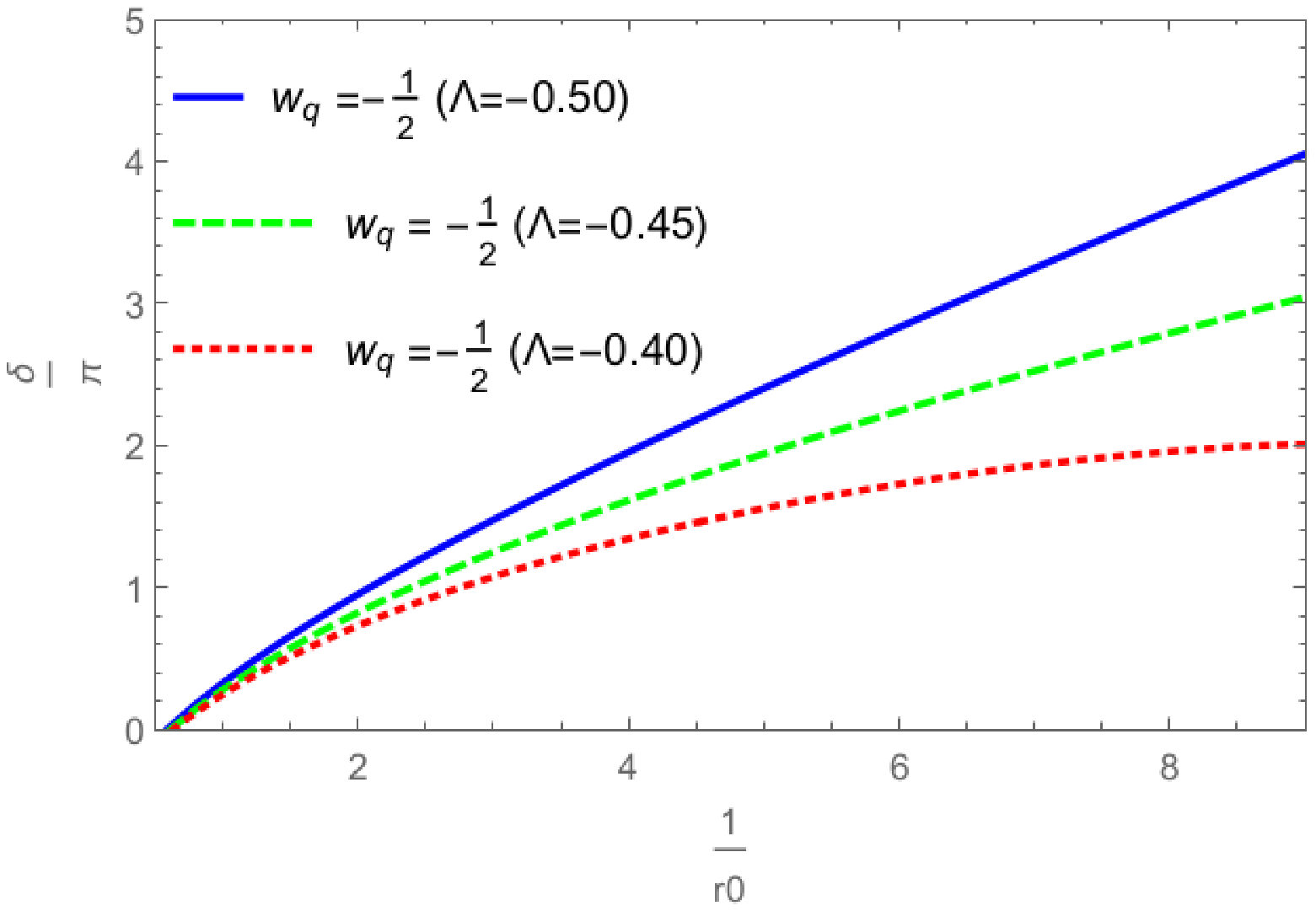}}
	\centering
	\caption{Pictorial representation of bending angle with inverse of the distance of closest approach for different values of cosmological constant (a) without quintessence and (b) with quintessence.} \label{v012}
\end{figure}

\begin{figure} [h]
	\begin{center}	
		\includegraphics[width=7cm,height=6cm]{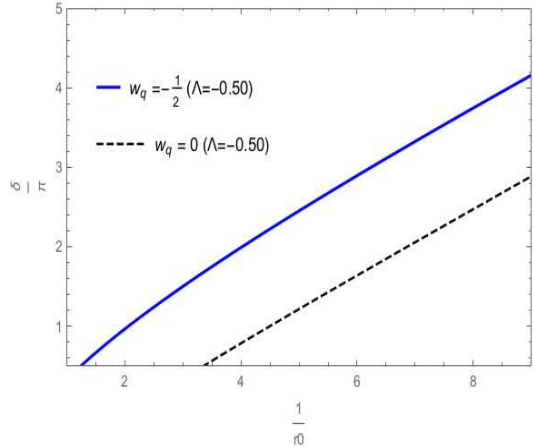}
	\end{center}
	\caption{The comparison of bending angle with and without quintessence for the same value of cosmological constant.} \label{voba}
\end{figure}

The variation of bending angle with inverse of the distance of closest approach for different values of cosmological constant depicted in \figurename{ \ref{v012}}. The bending angle for without quintessence (usual BTZ BH) increases as the $r_{0}$ decreases and the similar variation is observed for in presence of quintessence. In both the cases it is pointed out that the bending angle decreases with the significant increase of cosmological constant. It can be easily seen from \figurename{ \ref{v012}} that comparative behaviour of bending angle remain unaffected due to the presence of quintessence parameter. The comparison of bending angle with and without quintessence parameter is depicted in \figurename{ \ref{voba}}.
It is observed that the bending angle increases due to the presence of quintessence field i.e. the light goes closer to the BH in presence of  quintessence and more deflected than in the normal case.

\section{Summary and Conclusions} 

\noindent  In this paper, we have studied the equations of motion (i.e. geodesic equations) and the analytical solutions of null geodesic equations and timelike geodesic equations for EOS parameter $w_q=-1/2$ are obtained. In addition, we have calculated the effective potentials for light rays as well as test particles and depicted them graphically for different values of $w_q$ (i.e. $ -1/3, -2/3 $ and $ 0 $). It is observed that there is no significant variation within the above mentioned limit of $w_q$. However, it can be observed from \figurename{ \ref{epw}} that the radius of horizon decreases as the value of $w_q$ shifts towards  $ -1 $. It is also observed that no stable orbits are possible when $w_q$ lies between $0 < w_q < -2/3$.   Then, we have specified certain set of possible orbits for the massless test particles (i.e. light rays) and massive test particles with the use of these analytic solutions of geodesic equations and effective potentials. As shown in \figurename{ \ref{nuo}}, we have obtained EOs, TEOs and TBOs for light rays and in \figurename{ \ref{tlo}} TBOs and BOs are presented for massive particles. On the basis of number of positive zeros, we have  classified all these orbits in \tablename{ \ref{table}}.
We have compared the results obtained with the results of previously studied (2+1) dimensional BHs in GR \cite{cruz1994geodesic,Soroushfar:2015dfz} and it ts observed that the TOs and TEOs are also possible in usual BTZ BH. However, the BOs are obtained instead of CBOs due to presence of quintessential parameter  (for $ w_q = -1/2 $) for the present case.

Furthermore, we have obtained an exact expression of the bending angle for this BH spacetime with and without quintessence is derived in Eq. \ref{ba1} and Eq. \ref{ba2} respectively and studied its variation along with the distance of closest approach and cosmological constant. In both the cases, we have observed that the bending angle increases as the distance of closest approach decreases. We have also analyzed that in the absence of quintessence field, the bending angle gradually decreases as increase in cosmological constant from positive to negative region which in turn marks the unstable behaviour of deflection angle. It is noticeable that the bending angle increases in the presence of quintessence field and then goes to positive infinity which shows the stable behaviour of deflection angle. Such light paths of normal BHs and BHs in presence of the dynamical scalar fields have also been investigated in \cite{Azreg-Ainou:2017obt,AzregAinou:2012xv}. Based on this analysis one can emphasize that in the presence of phantom fields and quintessence field, the light rays are more deflected than in the normal case. The bending angle of light by BTZ BH surrounded by quintessence field and other results obtained by us reduces into usual BTZ BH for $w_q =0$ \cite{Kala:2020viz}. We have obtained that the presence of quintessence field increases the deflection angle more than usual case which is in close agreement with the observation made in \cite{Azreg-Ainou:2017obt,AzregAinou:2012xv}. With the results obtained here, one can argue to have some new insights of BHs in lower dimensional gravity as compared to usual BHs in four dimensions.

\section{Acknowledgments}
 The authors are grateful to the anonymous referee for the valuable comments
	and suggestions which helped us to improve the quality of this paper substantially. The authors SK and HN are thankful to the Uttarakhand State Council of Science and Technology (UCOST), Dehradun for financial assistance through R\&D grant number UCS\&T/RD-18/18-19/16038/4. The author HN and PS acknowledge the financial support provided by Science and Engineering Research Board (SERB), New Delhi through the grant number EMR/2017/000339. The authors (S. Kala, H. Nandan and P. Sharma) also acknowledges the facilities at ICARD, Gurukula Kangri (Deemed to be University), Haridwar, India. ME acknowledges that this work is supported by an NWU postdoctoral fellowship and the National Research Foundation (NRF) of South Africa (grant numbers 116657). 

\bibliographystyle{abbrv} 
\bibliography{sample} 

\begin{thebibliography}{10}

\bibitem{abramowitz1970handbook}
M.~Abramowitz and I.~A. Stegun.
\newblock {\em Handbook of mathematical functions with formulas, graphs, and
  mathematical tables}, volume~55.
\newblock US Government printing office, 1970.

\bibitem{AzregAinou:2012xv}
M.~Azreg-Ainou.
\newblock {Light paths of normal and phantom Einstein-Maxwell-dilaton black
  holes}.
\newblock {\em Phys. Rev. D}, 87(2):024012, 2013.

\bibitem{Azreg-Ainou:2014lua}
M.~Azreg-Aïnou.
\newblock {Charged de Sitter-like black holes: quintessence-dependent enthalpy
  and new extreme solutions}.
\newblock {\em Eur. Phys. J. C}, 75(1):34, 2015.

\bibitem{Azreg-Ainou:2017obt}
M.~Azreg-Aïnou, S.~Bahamonde, and M.~Jamil.
\newblock {Strong Gravitational Lensing by a Charged Kiselev Black Hole}.
\newblock {\em Eur. Phys. J. C}, 77(6):414, 2017.

\bibitem{AzregAinou:2012hy}
M.~Azreg-Aïnou and M.~E. Rodrigues.
\newblock {Thermodynamical, geometrical and Poincaré methods for charged black
  holes in presence of quintessence}.
\newblock {\em JHEP}, 09:146, 2013.

\bibitem{banados1999three}
M.~Banados.
\newblock Three-dimensional quantum geometry and black holes.
\newblock In {\em AIP Conference Proceedings}, volume 484, pages 147--169.
  American Institute of Physics, 1999.

\bibitem{banados1992black}
M.~Banados, C.~Teitelboim, and J.~Zanelli.
\newblock Black hole in three-dimensional spacetime.
\newblock {\em Physical Review Letters}, 69(13):1849, 1992.

\bibitem{bouali2019cosmological}
A.~Bouali, I.~Albarran, M.~Bouhmadi-L{\'o}pez, and T.~Ouali.
\newblock Cosmological constraints of phantom dark energy models.
\newblock {\em Physics of the Dark Universe}, 26:100391, 2019.

\bibitem{Bozza:2009yw}
V.~Bozza.
\newblock {Gravitational Lensing by Black Holes}.
\newblock {\em Gen. Rel. Grav.}, 42:2269--2300, 2010.

\bibitem{Carlip:1995qv}
S.~Carlip.
\newblock {The (2+1)-Dimensional black hole}.
\newblock {\em Class. Quant. Grav.}, 12:2853--2880, 1995.

\bibitem{Chatterjee:2019rym}
A.~K. Chatterjee, K.~Flathmann, H.~Nandan, and A.~Rudra.
\newblock {Analytic solutions of the geodesic equation for
  Reissner-Nordström--(anti--)de Sitter black holes surrounded by different
  kinds of regular and exotic matter fields}.
\newblock {\em Phys. Rev. D}, 100(2):024044, 2019.

\bibitem{Chen:2012mva}
S.~Chen, Q.~Pan, and J.~Jing.
\newblock {Holographic superconductors in quintessence AdS black hole
  spacetime}.
\newblock {\em Class. Quant. Grav.}, 30:145001, 2013.

\bibitem{cruz1994geodesic}
N.~Cruz, C.~Martinez, and L.~Pena.
\newblock Geodesic structure of the (2+ 1)-dimensional btz black hole.
\newblock {\em Classical and Quantum Gravity}, 11(11):2731, 1994.

\bibitem{deOliveira:2018weu}
J.~de~Oliveira and R.~D.~B. Fontana.
\newblock {Three-dimensional black holes with quintessence}.
\newblock {\em Phys. Rev.}, D98(4):044005, 2018.

\bibitem{eichler1982zeros}
M.~Eichler and D.~Zagier.
\newblock On the zeros of the weierstrass $\wp$-function.
\newblock {\em Mathematische Annalen}, 258(4):399--407, 1982.

\bibitem{Fernando:2014wma}
S.~Fernando.
\newblock {Cold, ultracold and Nariai black holes with quintessence}.
\newblock {\em Gen. Rel. Grav.}, 45:2053--2073, 2013.

\bibitem{Fernando:2014rsa}
S.~Fernando, S.~Meadows, and K.~Reis.
\newblock {Null trajectories and bending of light in charged black holes with
  quintessence}.
\newblock {\em Int. J. Theor. Phys.}, 54(10):3634--3653, 2015.

\bibitem{Guo:2004fq}
Z.-K. Guo, Y.-S. Piao, X.-M. Zhang, and Y.-Z. Zhang.
\newblock {Cosmological evolution of a quintom model of dark energy}.
\newblock {\em Phys. Lett. B}, 608:177--182, 2005.

\bibitem{hackmann2008analytic}
E.~Hackmann, V.~Kagramanova, J.~Kunz, and C.~L{\"a}mmerzahl.
\newblock Analytic solutions of the geodesic equation in higher dimensional
  static spherically symmetric spacetimes.
\newblock {\em Physical Review D}, 78(12):124018, 2008.

\bibitem{Hackmann:2008zz}
E.~Hackmann and C.~Lammerzahl.
\newblock {Geodesic equation in Schwarzschild- (anti-) de Sitter space-times:
  Analytical solutions and applications}.
\newblock {\em Phys. Rev.}, D78:024035, 2008.

\bibitem{Hackmann:2010zz}
E.~Hackmann, C.~Lammerzahl, V.~Kagramanova, and J.~Kunz.
\newblock {Analytical solution of the geodesic equation in Kerr-(anti) de
  Sitter space-times}.
\newblock {\em Phys. Rev. D}, 81:044020, 2010.

\bibitem{hagihara1930theory}
Y.~Hagihara.
\newblock Theory of the relativistic trajeetories in a gravitational field of
  schwarzschild.
\newblock {\em Japanese journal of astronomy and geophysics}, 8:67, 1930.

\bibitem{Hartmann:2010rr}
B.~Hartmann and P.~Sirimachan.
\newblock {Geodesic motion in the space-time of a cosmic string}.
\newblock {\em JHEP}, 08:110, 2010.

\bibitem{hashimoto2020imaging}
K.~Hashimoto, S.~Kinoshita, and K.~Murata.
\newblock Imaging black holes through the ads/cft correspondence.
\newblock {\em Physical Review D}, 101(6):066018, 2020.

\bibitem{Hyun:1997jv}
S.~Hyun.
\newblock {U duality between three-dimensional and higher dimensional black
  holes}.
\newblock {\em J. Korean Phys. Soc.}, 33:S532--S536, 1998.

\bibitem{hyun1999statistical}
S.~Hyun, W.~T. Kim, and J.~Lee.
\newblock Statistical entropy and ads-cft correspondence in btz black holes.
\newblock {\em Physical Review D}, 59(8):084020, 1999.

\bibitem{Ida:2000jh}
D.~Ida.
\newblock {No black hole theorem in three-dimensional gravity}.
\newblock {\em Phys. Rev. Lett.}, 85:3758--3760, 2000.

\bibitem{Iyer:2009wa}
S.~V. Iyer and E.~C. Hansen.
\newblock {Light's Bending Angle in the Equatorial Plane of a Kerr Black Hole}.
\newblock {\em Phys. Rev.}, D80:124023, 2009.

\bibitem{Kala:2020viz}
S.~Kala, H.~Nandan, and P.~Sharma.
\newblock {Deflection of Light Around a Rotating BTZ Black Hole}.
\newblock 10 2020.

\bibitem{Kazempour:2017gho}
S.~Kazempour and S.~Soroushfar.
\newblock {Investigation the geodesic motion of three dimensional rotating
  black holes}.
\newblock {\em Chin. J. Phys.}, 65:579--592, 2020.

\bibitem{Keeton:2005jd}
C.~R. Keeton and A.~Petters.
\newblock {Formalism for testing theories of gravity using lensing by compact
  objects. I. Static, spherically symmetric case}.
\newblock {\em Phys. Rev. D}, 72:104006, 2005.

\bibitem{kerr1963gravitational}
R.~P. Kerr.
\newblock Gravitational field of a spinning mass as an example of algebraically
  special metrics.
\newblock {\em Physical review letters}, 11(5):237, 1963.

\bibitem{kippenhahn1990stellar}
R.~Kippenhahn, A.~Weigert, and A.~Weiss.
\newblock {\em Stellar structure and evolution}.
\newblock Springer, 1990.

\bibitem{Kiselev:2002dx}
V.~V. Kiselev.
\newblock {Quintessence and black holes}.
\newblock {\em Class. Quant. Grav.}, 20:1187--1198, 2003.

\bibitem{Kunz:2006wc}
M.~Kunz and D.~Sapone.
\newblock {Crossing the Phantom Divide}.
\newblock {\em Phys. Rev. D}, 74:123503, 2006.

\bibitem{leibundgut2001cosmological}
B.~Leibundgut.
\newblock Cosmological implications from observations of type ia supernovae.
\newblock {\em Annual Review of Astronomy and Astrophysics}, 39(1):67--98,
  2001.

\bibitem{maldacena1999large}
J.~Maldacena.
\newblock The large-n limit of superconformal field theories and supergravity.
\newblock {\em International journal of theoretical physics}, 38(4):1113--1133,
  1999.

\bibitem{maoz2014observational}
D.~Maoz, F.~Mannucci, and G.~Nelemans.
\newblock Observational clues to the progenitors of type ia supernovae.
\newblock {\em Annual Review of Astronomy and Astrophysics}, 52:107--170, 2014.

\bibitem{Nozari:2020tks}
K.~Nozari and M.~Hajebrahimi.
\newblock {Geodesic Structure of the Quantum-Corrected Schwarzschild Black Hole
  Surrounded by Quintessence}.
\newblock 4 2020.

\bibitem{Peebles:2002gy}
P.~Peebles and B.~Ratra.
\newblock {The Cosmological Constant and Dark Energy}.
\newblock {\em Rev. Mod. Phys.}, 75:559--606, 2003.

\bibitem{riess1998observational}
A.~G. Riess, A.~V. Filippenko, P.~Challis, A.~Clocchiatti, A.~Diercks, P.~M.
  Garnavich, R.~L. Gilliland, C.~J. Hogan, S.~Jha, R.~P. Kirshner, et~al.
\newblock Observational evidence from supernovae for an accelerating universe
  and a cosmological constant.
\newblock {\em The Astronomical Journal}, 116(3):1009, 1998.

\bibitem{Sfetsos:1997xs}
K.~Sfetsos and K.~Skenderis.
\newblock {Microscopic derivation of the Bekenstein-Hawking entropy formula for
  nonextremal black holes}.
\newblock {\em Nucl. Phys. B}, 517:179--204, 1998.

\bibitem{Sharma:2019qxd}
P.~Sharma, H.~Nandan, R.~Gannouji, R.~Uniyal, and A.~Abebe.
\newblock {Deflection of Light by a Rotating Black Hole Surrounded by
  Quintessence}.
\newblock 11 2019.

\bibitem{Soroushfar:2015dfz}
S.~Soroushfar, R.~Saffari, and A.~Jafari.
\newblock {Study of geodesic motion in a ( 2+1 )-dimensional charged BTZ black
  hole}.
\newblock {\em Phys. Rev. D}, 93(10):104037, 2016.

\bibitem{Soroushfar:2015wqa}
S.~Soroushfar, R.~Saffari, J.~Kunz, and C.~Lämmerzahl.
\newblock {Analytical solutions of the geodesic equation in the spacetime of a
  black hole in f(R) gravity}.
\newblock {\em Phys. Rev.}, D92(4):044010, 2015.

\bibitem{Steinhardt:2003st}
P.~J. Steinhardt.
\newblock {A quintessential introduction to dark energy}.
\newblock {\em Phil. Trans. Roy. Soc. Lond.}, A361:2497--2513, 2003.

\bibitem{sullivan2006rates}
M.~Sullivan, D.~Le~Borgne, C.~Pritchet, A.~Hodsman, J.~Neill, D.~Howell,
  R.~Carlberg, P.~Astier, E.~Aubourg, D.~Balam, et~al.
\newblock Rates and properties of type ia supernovae as a function of mass and
  star formation in their host galaxies.
\newblock {\em The Astrophysical Journal}, 648(2):868, 2006.

\bibitem{Thomas:2012zzc}
B.~B. Thomas, M.~Saleh, and T.~C. Kofane.
\newblock {Thermodynamics and phase transition of the Reissner-Nordstroem black
  hole surrounded by quintessence}.
\newblock {\em Gen. Rel. Grav.}, 44:2181--2189, 2012.

\bibitem{Uniyal:2018ngj}
R.~Uniyal, H.~Nandan, and P.~Jetzer.
\newblock {Bending angle of light in equatorial plane of Kerr--Sen Black Hole}.
\newblock {\em Phys. Lett. B}, 782:185--192, 2018.

\bibitem{Virbhadra:1999nm}
K.~S. Virbhadra and G.~F.~R. Ellis.
\newblock {Schwarzschild black hole lensing}.
\newblock {\em Phys. Rev.}, D62:084003, 2000.

\bibitem{Wang:2016lxa}
B.~Wang, E.~Abdalla, F.~Atrio-Barandela, and D.~Pavon.
\newblock {Dark Matter and Dark Energy Interactions: Theoretical Challenges,
  Cosmological Implications and Observational Signatures}.
\newblock {\em Rept. Prog. Phys.}, 79(9):096901, 2016.

\bibitem{whittaker2020course}
E.~T. Whittaker and G.~N. Watson.
\newblock {\em A course of modern analysis}.
\newblock Dover Publications, 2020.

\bibitem{xia2006features}
J.-Q. Xia, G.-B. Zhao, H.~Li, B.~Feng, and X.~Zhang.
\newblock Features in the dark energy equation of state and modulations in the
  hubble diagram.
\newblock {\em Physical Review D}, 74(8):083521, 2006.

\bibitem{Yang:2009zzl}
R.-J. Yang and X.-T. Gao.
\newblock {Observational constraints on purely kinetic k-essence dark energy
  models}.
\newblock {\em Chin. Phys. Lett.}, 26:089501, 2009.

\bibitem{zehavi1999evidence}
I.~Zehavi and A.~Dekel.
\newblock Evidence for a positive cosmological constant from flows of galaxies
  and distant supernovae.
\newblock {\em Nature}, 401(6750):252--254, 1999.

\bibitem{Zhou:2007xp}
S.-Y. Zhou.
\newblock {A New Approach to Quintessence and Solution of Multiple Attractors}.
\newblock {\em Phys. Lett. B}, 660:7--12, 2008.

\end{thebibliography}
\end{document}